\documentclass[12pt,preprint]{aastex}

\usepackage{amsmath}
\usepackage{natbib}
\usepackage{graphicx}
\usepackage{color}
\usepackage[breaklinks,colorlinks,citecolor=blue]{hyperref}
\usepackage{epsfig}
\usepackage{epstopdf}

\usepackage{mathrsfs}

\begin{document}
\title{Kerr-Anti-De-Sitter/De-Sitter Black Hole In Perfect Fluid Dark Matter Background}

\author{
  Zhaoyi Xu,\altaffilmark{1,2,3,4}
  Xian Hou,\altaffilmark{1,3,4}
  Jiancheng Wang \altaffilmark{1,2,3,4}
 }

\altaffiltext{1}{Yunnan Observatories, Chinese Academy of Sciences, 396 Yangfangwang, Guandu District, Kunming, 650216, P. R. China; {\tt zyxu88@ynao.ac.cn,xianhou.astro@gmail.com,jcwang@ynao.ac.cn}}
\altaffiltext{2}{University of Chinese Academy of Sciences, Beijing, 100049, P. R. China}
\altaffiltext{3}{Key Laboratory for the Structure and Evolution of Celestial Objects, Chinese Academy of Sciences, 396 Yangfangwang, Guandu District, Kunming, 650216, P. R. China}
\altaffiltext{4}{Center for Astronomical Mega-Science, Chinese Academy of Sciences, 20A Datun Road, Chaoyang District, Beijing, 100012, P. R. China}

\shorttitle{Kerr-Anti-De-Sitter/De-Sitter Black Hole In Perfect Fluid Dark Matter Background}
\shortauthors{Z Y. Xu et al.}

\begin{abstract}
We obtain the Kerr-anti-de-sitter (Kerr-AdS) and Kerr-de-sitter (Kerr-dS) black hole (BH) solutions to the Einstein field equation in the perfect fluid dark matter background using the Newman-Janis method and Mathematica package. We discuss in detail the black hole properties and obtain the following main results: (i) From the horizon equation $g_{rr}=0$, we derive the relation between the perfect fluid dark matter parameter $\alpha$ and the cosmological constant $\Lambda$ when the cosmological horizon $r_{\Lambda}$ exists. For $\Lambda=0$, we find that $\alpha$ is in the range $0<\alpha<2M$ for $\alpha>0$ and $-7.18M<\alpha<0$ for $\alpha<0$. For positive cosmological constant $\Lambda$ (Kerr-AdS BH), $\alpha_{max}$ decreases if $\alpha>0$, and $\alpha_{min}$ increases if $\alpha<0$. For negative cosmological constant $-\Lambda$ (Kerr-dS BH), $\alpha_{max}$ increases if $\alpha>0$ and $\alpha_{min}$ decreases if $\alpha<0$; (ii) An ergosphere exists between the event horizon and the outer static limit surface. The size of the ergosphere evolves oppositely for $\alpha>0$ and $\alpha<0$, while decreasing with the increasing $\mid\alpha\mid$. When there is sufficient dark matter around the black hole, the black hole spacetime changes remarkably; (iii) The singularity of these black holes is the same as that of rotational black holes. In addition, we study the geodesic motion using the Hamilton-Jacobi formalism and find that when $\alpha$ is in the above ranges for $\Lambda=0$, stable orbits exist. Furthermore, the rotational velocity of the black hole in the equatorial plane has different behaviour for different $\alpha$ and the black hole spin $a$. It is asymptotically flat and independent of $\alpha$ if $\alpha>0$ while is asymptotically flat only when $\alpha$ is close to zero if $\alpha<0$. We anticipate that Kerr-Ads/dS black holes could exist in the universe and our future work will focus on the observational effects of the perfect fluid dark matter on these black holes.
\end{abstract}


\section{INTRODUCTION}
Recent cosmological observations reveal that the universe is full of dark matter (DM) and dark energy which place the universe in accelerated expansion \citep{2006IJMPD..15.1753C,2011ApJS..192...18K,2003RvMP...75..559P,1999ApJ...517..565P,2004ApJ...607..665R}. If the universe is filled with a perfect dark-energy like fluid, the equation of state satisfies $\omega<-1/3$, where the state parameter is $\omega=p/\rho$, with $p$ and $\rho$ the energy density and matter pressure, respectively. Observations indicate that the equation of state is very close to the cosmological constant case $\omega=-1$ \citep{2011ApJS..192...18K,2004MNRAS.354..275A}. Many dark energy models have been studied in the literature, for example, the quintessence dark energy model with $-1<\omega<-1/3$ \citep{2006IJMPD..15.1753C}, the phantom dark energy models with $\omega<-1$ \citep{2002PhLB..545...23C,2003PhRvL..91g1301C} and the quitom dark energy model with $\omega$ crosses $-1$ \citep{2010PhR...493....1C}. The phantom dark energy models violate the dominant energy condition, and could cause a ``Big Rip'' singularity \citep{2003PhRvL..91g1301C,2004PhRvD..70l3529N}.

Observation of the cosmic microwave background and of large-scale structure show that the geometry of our universe is nearly spatially flat \citep{2011ApJS..192...18K,2003ApJS..148..175S}. According to the standard model of cosmology, our current universe contains around 68$\%$ dark energy, around 28$\%$ DM and less than 4$\%$ baryonic matter. Here, DM is non-baryonic and non-luminous. Weakly interacting massive particles (WIMPs) are a kind of cold DM (CDM), for which the equation of state $\omega$ is close to $0$ with dust-like properties. However, observations revealed that the collisionless CDM paradigm breaks down on small scales. These small-scale problems include the core-cusp problem, rotation curve diversity problem, missing satellites problem, too-big-to-fail problem and baryon feedback problem \citep[See a recent review,][]{2018PhR...730....1T}. In order to solve these problems, some alternative models have been proposed, such as Warm Dark Matter and Fuzzy Dark Matter. On the other hand, these dark matter models could be regarded as perfect fluid DM. We are interested in how black hole solutions depend on the perfect fluid DM parameters.

The influence of quintessence dark energy on spherical symmetric black holes has been obtained by \cite{2003CQGra..20.1187K}, and the properties of black holes surrounded by quintessence matter have been discussed extensively in the literature \citep{2008CoTPh..50..101C,2013PhRvD..88j4007D,2012GReGr..44.1857F,2009CoTPh..52..974G,2015EPJC...75...24J,2014Ap&SS.349..865K,
2008arXiv0805.4554K,2010GReGr..42.1719L,2015Ap&SS.357..112M,2015MPLA...3050049M,2015Ap&SS.357....8M,2012GReGr..44.2181T,2015GReGr..47...16U,2009GReGr..41.1249V,2006PhRvD..74l6002K,2014arXiv1405.3931P,2014arXiv1405.6113P,2014IJMPA..2950164S,Singh2014,2016arXiv161005454X,2017arXiv171104542X}. For rotational black holes, the Kerr-Newman-AdS black hole solutions for quintessence dark energy have also been obtained recently \citep{2016EPJC...76..222G,2017EPJP..132...98T,2017PhRvD..95f4015X}. On the other hand, spherically symmetric black hole solutions surrounded by perfect fluid DM have been obtained and invoked to explain the asymptotically flat rotation velocity in spiral galaxies \citep{2005CQGra..22..541K,2004CQGra..21.3323K,2003gr.qc.....3031K,2012PhRvD..86l3015L}.
In this work, we first generalize the spherical symmetric black hole solution with perfect fluid DM to Kerr-like black holes using the Newman-Janis method. Then we extend this Kerr-like solution to Kerr-anti-de-sitter (Kerr-AdS) and Kerr-de-sitter (Kerr-dS) black holes to include a cosmological constant.

The paper is organized as follows. In Section \ref{spheric}, we introduce the solution of the spherically symmetric black hole surrounded by perfect fluid DM. In Section \ref{kerrads}, we generalize this solution to Kerr-AdS/dS black holes with the Newman-Janis method. In Section \ref{properties}, we study the properties of Kerr-AdS/dS black holes in perfect fluid DM, including the horizons, the stationary limit surfaces and the singularity. In Section \ref{geodesic}, we derive the geodesic structure in the equatorial plane. Finally we give our summary in Section \ref{summary}.

\section{Spherically symmetric black hole solution in perfect fluid dark matter}
\label{spheric}
When the black hole is surrounded by DM, the DM field couples with the gravity. The action is given by
\begin{equation}
S=\int d^{4}x\sqrt{-g}(\dfrac{1}{16\pi G}R+\mathcal{L}_{DM}),
\label{Action1}
\end{equation}
where $G$ is Newton's constant and $\mathcal{L}_{DM}$ is the DM Lagrangian density. In Eq.(\ref{Action1}) we have not included the interaction between DM and ordinary matter fields. With the variational approach, we obtain the Einstein field equation
\begin{equation}
R_{\mu\nu}-\dfrac{1}{2}g_{\mu\nu}R=-8\pi G (\bar{T_{\mu\nu}}+T_{\mu\nu}(DM))=-8\pi G T_{\mu\nu},
\label{einstein equation 1}
\end{equation}
where $\bar{T_{\mu\nu}}$ is the energy-momentum tensor of the ordinary matter and $T_{\mu\nu}(DM)$ is the energy-momentum tensor of DM. In the case of black holes surrounded by DM, assuming that DM is a kind of perfect fluid, the energy-momentum tensor can then be written as $T^{\mu}_{~~\mu}=diag[-\rho,p,p,p]$ (where $T^{\mu}_{~~\mu}=g^{\mu\nu}T_{\mu\nu}$). In addition, in the simplest case, we assume $\delta-1=p/\rho$, where $\delta$ is a constant \citep{2012PhRvD..86l3015L}. We refer to such DM as ``perfect fluid DM" in this work.

For the spherically symmetric spacetime metric in perfect fluid DM, the black hole solution is \citep{2003gr.qc.....3031K,2012PhRvD..86l3015L}
\begin{equation}
ds^{2}=-f(r)dt^{2}+f^{-1}(r)dr^{2}+r^{2}(d\theta^{2}+sin^{2}\theta d\phi^{2}),
\label{solution1}
\end{equation}
where
\begin{equation}
f(r)=1-\dfrac{2M}{r}+\dfrac{\alpha}{r}ln(\dfrac{r}{\mid\alpha \mid}),
\label{solution2}
\end{equation}
$M$ is the black hole mass and $\alpha$ is a parameter describing the intensity of the perfect fluid DM. This solution corresponds to a specific case of the general solution in \cite{2003gr.qc.....3031K} and \cite{2012PhRvD..86l3015L}.
It is interesting to note that this black hole solution implies that the rotational velocity is asymptotically flat in the equatorial plane, which could explain the observed rotation curves in spiral galaxies \citep{2003gr.qc.....3031K,2012PhRvD..86l3015L}. 

\section{KERR-AdS/dS BLACK HOLE SOLUTION IN PERFECT FLUID DARK MATTER}
\label{kerrads}

\subsection{Kerr-like black hole in perfect fluid dark matter}
Using the Newman-Janis method, we now generalize the spherically symmetric black hole solution in perfect fluid DM to Kerr-like rotational black holes. The Newman-Janis method has been introduced in many articles \citep{1965JMP.....6..915N,2014PhRvD..90f4041A,2017EPJP..132...98T,2017PhRvD..95f4015X}. As the first step of the Newman-Janis method, we transform the spherically symmetric black hole spacetime metric (Eq. \ref{solution1}) from Boyer-Lindquist (BL) coordinates $(t,r,\theta,\phi)$ to Eddington-Finkelstein (EF) coordinates $(u,r,\theta,\phi)$ through
\begin{equation}
du=dt-\dfrac{dr}{1-\dfrac{2M}{r}+\dfrac{\alpha}{r}ln(\dfrac{r}{\mid\alpha \mid})}.
\label{NJ1}
\end{equation}
Eq. (\ref{solution1}) can then be rewritten as
\begin{equation}
ds^{2}=-(1-\dfrac{2M}{r}+\dfrac{\alpha}{r}ln(\dfrac{r}{\mid\alpha \mid}))du^{2}-2dudr+r^{2}d\Omega^{2}.
\label{6}
\end{equation}
The non-zero components of the inverse spacetime metric in the null (EF) coordinates derived from Eq. (\ref{6}) are
\begin{equation}
g^{rr}=1-\dfrac{2M}{r}+\dfrac{\alpha}{r}ln(\dfrac{r}{\mid\alpha \mid}),~~~    g^{\theta\theta}=\dfrac{1}{r^{2}}, $$$$  g^{\phi\phi}=\dfrac{1}{r^{2}sin^{2}\theta}, ~~~ g^{ur}=g^{ru}=-1.
\label{7}
\end{equation}
From the definition of the null tetrad, the metric matrix is given by
\begin{equation}
g^{\mu\nu}=-l^{\mu}n^{\nu}-l^{\nu}n^{\mu}+m^{\mu}\overline{m}^{\nu}+m^{\nu}\overline{m}^{\mu}.
\label{8}
\end{equation}
The null vectors of the null tetrad satisfy the relations $l_{\mu}l^{\mu}=n_{\mu}n^{\mu}=m_{\mu}m^{\mu}=l_{\mu}m^{\mu}=n_{\mu}m^{\mu}=0, l_{\mu}n^{\mu}=-m_{\mu}\overline{m}^{\mu}=1$
and the corresponding components are
\begin{equation}
l^{\mu}=\delta^{\mu}_{r},$$$$
n^{\mu}=\delta^{\mu}_{0}-\dfrac{1}{2}(1-\dfrac{2M}{r}+\dfrac{\alpha}{r}ln(\dfrac{r}{\mid\alpha \mid}))\delta^{\mu}_{r},$$$$
m^{\mu}=\dfrac{1}{\sqrt{2}r}\delta^{\mu}_{\theta}+\dfrac{i}{\sqrt{2}r sin\theta}\delta^{\mu}_{\phi},$$$$
\overline{m}^{\mu}=\dfrac{1}{\sqrt{2}r}\delta^{\mu}_{\theta}-\dfrac{i}{\sqrt{2}r sin\theta}\delta^{\mu}_{\phi}.
\label{9}
\end{equation}
Working with these null vectors, we perform the coordinate transformation in the $(u,r)$ space as
\begin{equation}
u\longrightarrow u-iacos\theta, $$$$
r\longrightarrow r-iacos\theta.
\label{10}
\end{equation}
At the same time, we change $f(r)$ to $F(r,a,\theta)$ and $g(r)$ to $ G(r,a,\theta)$ (here $g(r)=f(r)$), and define $\Sigma^{2}=r^{2}+a^{2}cos^{2}\theta$. In the $(u,r)$ space, the null vectors are therefore given by
\begin{equation}
l^{\mu}=\delta^{\mu}_{r},~~~~n^{\mu}=\sqrt{\dfrac{G}{F}}\delta^{\mu}_{0}-\dfrac{1}{2}F\delta^{\mu}_{r},$$$$
m^{\mu}=\dfrac{1}{\sqrt{2}\Sigma}(\delta^{\mu}_{\theta}+ia sin\theta(\delta^{\mu}_{0}-\delta^{\mu}_{r})+\dfrac{i}{sin\theta}\delta^{\mu}_{\phi}),$$$$
\overline{m}^{\mu}=\dfrac{1}{\sqrt{2}\Sigma}(\delta^{\mu}_{\theta}-ia sin\theta(\delta^{\mu}_{0}-\delta^{\mu}_{r})-\dfrac{i}{sin\theta}\delta^{\mu}_{\phi}).
\label{11}
\end{equation}
From the definition of the null tetrades (Eq. \ref{8}), we obtain the spacetime metric tensor $g^{\mu\nu}$ as
\begin{equation}
g^{uu}=\dfrac{a^{2}sin^{2}\theta}{\Sigma^{2}},~~~~~~g^{rr}=G+\dfrac{a^{2}sin^{2}\theta}{\Sigma^{2}},$$$$
g^{\theta\theta}=\dfrac{1}{\Sigma^{2}},~~~~~~g^{\phi\phi}=\dfrac{1}{\Sigma^{2}sin^{2}\theta},$$$$
g^{ur}=g^{ru}=-\sqrt{\dfrac{G}{F}}-\dfrac{a^{2}sin^{2}\theta}{\Sigma^{2}},$$$$
g^{u\phi}=g^{\phi u}=\dfrac{a}{\Sigma^{2}},~~~~~~g^{r\phi}=g^{\phi r}=-\dfrac{a}{\Sigma^{2}}.
\label{12}
\end{equation}
The covariant metric tensor in the EF coordinates of $(u,r,\theta,\phi)$ are given by
\begin{equation}
g_{uu}=-F,~~~~~~g_{\theta\theta}=\Sigma^{2},~~~~~~g_{ur}=g_{ru}=-\sqrt{\dfrac{G}{F}},$$$$
g_{\phi\phi}=sin^{2}\theta(\Sigma^{2}+a^{2}(2\sqrt{\dfrac{F}{G}}-F)sin^{2}\theta),$$$$
g_{u\phi}=g_{\phi u}=a(F-\sqrt{\dfrac{F}{G}})sin^{2}\theta,~~~~~~g_{r\phi}=g_{\phi r}=a sin^{2}\theta\sqrt{\dfrac{F}{G}}.
\label{13}
\end{equation}
Finally, we perform the coordinate transformation between the EF and BL coordinates as
\begin{equation}
du=dt+\lambda(r)dr,~~~~d\phi=d\phi+h(r)dr,
\label{14}
\end{equation}
where
\begin{equation}
\lambda(r)=-\dfrac{r^{2}+a^{2}}{r^{2}g(r)+a^{2}},~~~h(r)=-\dfrac{a}{r^{2}g(r)+a^{2}},$$$$
F(r,\theta)=G(r,\theta)=\dfrac{r^{2}g(r)+a^{2}cos^{2}\theta}{\Sigma^{2}}.
\label{15}
\end{equation}
The Kerr-like black hole solution with the perfect fluid DM in the BL coordinates of $(t,r,\theta,\phi)$ are therefore given by

\begin{equation}
ds^{2}=-(1-\dfrac{2Mr-\alpha r ln(\dfrac{r}{\mid\alpha\mid})}{\Sigma^{2}})dt^{2}+\dfrac{\Sigma^{2}}{\Delta_{r}}dr^{2}-\dfrac{2a sin^{2}\theta(2Mr-\alpha r ln(\dfrac{r}{\mid\alpha\mid}))}{\Sigma^{2}}d\phi dt+\Sigma^{2}d\theta^{2}$$$$
+sin^{2}\theta (r^{2}+a^{2}+a^{2}sin^{2}\theta\dfrac{2Mr-\alpha r ln(\dfrac{r}{\mid\alpha\mid})}{\Sigma^{2}})d\phi^{2},
\label{solution10}
\end{equation}
where
\begin{equation}
\Delta_{r}=r^{2}-2Mr+a^{2}+\alpha r ln(\dfrac{r}{\mid\alpha\mid}).
\label{solution11}
\end{equation}
If the perfect fluid DM is absent $(\alpha=0)$, the above solution reduces to that of Kerr black holes.

\subsection{Kerr-AdS/dS black hole in perfect fluid dark matter}
We now extend the Kerr-like black hole solutions obtained above to the Kerr-AdS/dS black hole by including the cosmological constant $\pm\Lambda$. In the perfect fluid DM background, we rewrite the Kerr-like black hole metric (Eq. 16) as
\begin{equation}
ds^{2}=\dfrac{\Sigma^{2}}{\Delta_{r}}dr^{2}+\Sigma^{2}d\theta^{2}+\dfrac{sin^{2}\theta}{\Sigma^{2}}(adt-(r^{2}+a^{2})d\phi)^{2}-\dfrac{\Delta_{r}}{\Sigma^{2}}(dt-a sin^{2}d\phi)^{2}.
\label{AdS1}
\end{equation}
From the Einstein tensor formula of $G_{\mu\nu}=R_{\mu\nu}-\dfrac{1}{2}Rg_{\mu\nu}$ for the Kerr-like spacetime metric, we can obtain its expression through calculations utilizing the Mathematica package. The Einstein field equation taking into account the cosmological constant $\pm\Lambda$ is
\begin{equation}
\widetilde{G}_{\mu\nu}=R_{\mu\nu}-\dfrac{1}{2}R g_{\mu\nu}\pm\Lambda g_{\mu\nu}=8\pi T_{\mu\nu}.
\label{AdS3}
\end{equation}

Because the Newman-Jains method does not include the cosmological constant, we use other methods to obtain the solution with the cosmological constant, as introduced in \cite{2017PhRvD..95f4015X}. First, we guess the solution to the Einstein field equation with cosmological constant as
\begin{equation}
ds^{2}=\dfrac{\Sigma^{2}}{\Delta_{r}}dr^{2}+\dfrac{\Sigma^{2}}{\Delta_{\theta}}d\theta^{2}+\dfrac{\Delta_{\theta}sin^{2}\theta}{\Sigma^{2}}(a\dfrac{dt}{\Xi}-(r^{2}+a^{2})\dfrac{d\phi}{\Xi})^{2}-\dfrac{\Delta_{r}}{\Sigma^{2}}(\dfrac{dt}{\Xi}-a sin^{2}\dfrac{d\phi}{\Xi})^{2},
\label{AdS5}
\end{equation}
where
\begin{equation}
\Delta_{r}=r^{2}-2Mr+a^{2}\mp\dfrac{\Lambda}{3}r^{2}(r^{2}+a^{2})-\alpha r ln(\dfrac{r}{\mid\alpha\mid}),$$$$
\Delta_{\theta}=1\pm\dfrac{\Lambda}{3}a^{2}cos^{2}\theta~~~~~~~~\Xi=1\pm\dfrac{\Lambda}{3}a^{2}.
\label{AdS6}
\end{equation}

To verify that this black hole metric is the solution to the Einstein field equation (Eq. \ref{AdS3}), we insert Eq. (\ref{AdS5}) into Eq. (\ref{AdS3})
and use the Mathematica package to obtain the non-zero Einstein tensor as
\begin{equation}
\widetilde{G}_{tt}=\dfrac{\alpha}{\Sigma^{6}r}[r^{4}-2r^{3}M-r^{3}\alpha ln(\dfrac{r}{\mid\alpha\mid})+a^{2}r^{2}-a^{4}sin^{2}\theta cos^{2}\theta]+\dfrac{\alpha a^{2}sin^{2}\theta}{2\Sigma^{4}r},
$$$$
\widetilde{G}_{rr}=\dfrac{-\alpha r}{\Sigma^{2}\Delta_{r}},$$$$
\widetilde{G}_{t\phi}=\dfrac{\alpha a sin^{2}\theta}{r\Sigma^{6}}[(r^{2}+a^{2})(a^{2}cos^{2}\theta-r^{2})]+\dfrac{\alpha a^{2}sin^{2}\theta(r^{2}+a^{2})}{2r\Sigma^{4}},
$$$$
\widetilde{G}_{\theta\theta}=-\dfrac{\alpha a^{2}cos^{2}\theta}{r\Sigma^{2}}+\dfrac{\alpha}{2r},$$$$
\widetilde{G}_{\phi\phi}=-\dfrac{\alpha a^{2}sin^{2}\theta}{2r\Sigma^{6}}[(r^{2}+a^{2})(a^{2}+(2r^{2}+a^{2})cos2\theta)+2r^{3}M sin^{2}\theta+\alpha r^{3}M sin^{2}\theta ln(\dfrac{r}{\mid\alpha\mid}))]+$$$$
\dfrac{\alpha sin^{2}\theta(r^{2}+a^{2})^{2}}{2r\Sigma^{4}}.
\label{AdS7}
\end{equation}
We find the non-zero Einstein tensor $\widetilde{G}_{\mu\nu}$ (Eq. 22) with cosmological constant indeed equals the non-zero Einstein tensor $G_{\mu\nu}$ of the Kerr-like black hole spacetime. Therefore the Kerr-AdS/dS black hole spacetime (Eq. \ref{AdS5}) satisfies the Einstein field equation (Eq. \ref{AdS3}) in the perfect fluid DM with cosmological constant.

In the following, we give the form of the energy-momentum tensor of the Einstein field equation (Eq. \ref{AdS3}). From the space-time metric (Eq. \ref{AdS5}), we obtain the standard orthogonal basis of the Kerr-AdS/dS black hole metric in perfect fluid DM as
\begin{equation}
e^{\mu}_{t}=\dfrac{1}{\sqrt{\Xi^{2}\Sigma^{2}\Delta_{r}}}(r^{2}+a^{2},0,0,a),$$$$
e^{\mu}_{r}=\dfrac{\sqrt{\Delta_{r}}}{\sqrt{\Sigma^{2}}}(0,1,0,0),$$$$
e^{\mu}_{\theta}=\dfrac{\sqrt{\Delta_{\theta}}}{\sqrt{\Sigma^{2}}}(0,0,1,0),$$$$
e^{\mu}_{\phi}=-\dfrac{1}{\sqrt{\Xi^{2}\Sigma^{2}sin^{2}\theta}}(r^{2}+a^{2},0,0,a).
\label{EM1}
\end{equation}
From this standard orthogonal basis, we derive the relation between the non-zero components of the energy-momentum tensor and the Einstein tensor as
\begin{equation}
E=-\dfrac{1}{8\pi}e^{\mu}_{t}e^{\nu}_{t}G_{\mu\nu}=-\dfrac{1}{8\pi}e^{\mu}_{t}e^{\nu}_{t}(R_{\mu\nu}-\dfrac{1}{2}Rg_{\mu\nu}),$$$$
P_{r}=\dfrac{1}{8\pi}e^{\mu}_{r}e^{\nu}_{r}G_{\mu\nu}=\dfrac{1}{8\pi}g^{rr}(R_{rr}-\dfrac{1}{2}Rg_{rr}),$$$$
P_{\theta}=\dfrac{1}{8\pi}e^{\mu}_{\theta}e^{\nu}_{\theta}G_{\mu\nu}=\dfrac{1}{8\pi}g^{\theta\theta}(R_{\theta\theta}-\dfrac{1}{2}Rg_{\theta\theta}),$$$$
P_{\phi}=-\dfrac{1}{8\pi}e^{\mu}_{\phi}e^{\nu}_{\phi}G_{\mu\nu}=-\dfrac{1}{8\pi}e^{\mu}_{\phi}e^{\nu}_{\phi}(R_{\mu\nu}-\dfrac{1}{2}Rg_{\mu\nu}).
\label{EM2}
\end{equation}
Combining Eqs. (20-24), we obtain the expression of the energy-momentum tensor for the Kerr-AdS/dS black hole surrounded by perfect fluid DM as
\begin{equation}
E=-P_{r}=\dfrac{\alpha r}{8\pi(r^{2}+a^{2}cos^{2}\theta)^{2}},$$$$
P_{\theta}=P_{\phi}=-P_{r}-\dfrac{\alpha}{16\pi r(r^{2}+a^{2}cos^{2}\theta)}=\dfrac{\alpha}{8\pi(r^{2}+a^{2}cos^{2}\theta)^{2}}(r-\dfrac{r^{2}+a^{2}cos^{2}\theta}{2r}).
\label{EM3}
\end{equation}

\section{KERR-AdS/dS BLACK HOLE PROPERTIES IN PERFECT FLUID DARK MATTER}
\label{properties}
\subsection{Horizons}
Stationary rotational black holes always possess the horizons which determine the black hole properties, such as thermodynamic quantities, classical geometric features, etc.
In order to obtain the horizons for Kerr-AdS/dS black holes, we give the horizon definition as
\begin{equation}
g^{\mu\nu}\dfrac{\partial f}{\partial x^{\mu}}\dfrac{\partial f}{\partial x^{\nu}}=0,
\label{EHD1}
\end{equation}
where $f(x^{\mu})=f(t,r,\theta,\phi)=0$ is a three-dimensional hypersurface in four-dimensional spacetime. When the black hole is axisymmetric and stationary, $f$ reduces to $f(r,\theta)$ for the black hole spacetime (Eq. \ref{AdS5}), and Eq. \ref{EHD1} reduces to
\begin{equation}
g^{rr}(\dfrac{\partial f}{\partial r})^{2}+g^{\theta\theta}(\dfrac{\partial f}{\partial \theta})^{2}=0.
\label{EHD2}
\end{equation}
From calculations, we find that the Kerr-AdS/dS black hole surrounded by the perfect fluid DM also has a horizon defined by
\begin{equation}
\Delta_{r}=r^{2}-2Mr+a^{2}\mp\dfrac{\Lambda}{3}r^{2}(r^{2}+a^{2})-\alpha r ln(\dfrac{r}{\mid\alpha\mid})=0,
\label{E1}
\end{equation}
which depends on $a,\Lambda$ and $\alpha$. After some algebra, Eq. (\ref{E1}) becomes
\begin{equation}
\pm\dfrac{\Lambda}{3}r^{4}+(\pm\dfrac{\Lambda}{3}a^{2}-1)r^{2}+2Mr-a^{2}=-\alpha r ln(\dfrac{r}{\mid\alpha\mid}).
\label{E2}
\end{equation}
By analyzing the number of horizons in Eq. (\ref{E2}), we can constrain the value of the DM parameter $\alpha$. Here we consider three cases corresponding to $\Lambda=0$ (Kerr-like), $\Lambda$ (Kerr-AdS) and $-\Lambda$ (Kerr-dS). Since perfect fluid DM does not change the number of horizons, there are two horizons for $\Lambda=0$ (the inner horizon and the event horizon), three horizons for positive $\Lambda$ (the inner horizon, the event horizon and the cosmological horizon) and two horizons for negative $\Lambda$ (the inner horizon and the event horizon).\\

Case I: $\Lambda=0$ (Kerr-like)

The horizon equation (Eq. \ref{E2}) becomes
\begin{equation}
r^{2}-2Mr+a^{2}=\alpha r ln(\dfrac{r}{\mid\alpha\mid}),
\label{E3}
\end{equation}
which has two horizons, the Cauchy (inner) horizon $r_{-}$ and the event horizon $r_{+}$. DM changes the size of two horizons. Since DM does not produce new horizons, Eq. (\ref{E3}) has two roots and one extreme value point. Taking the derivative of Eq. (\ref{E3}) with respect to $r$ results in
\begin{equation}
r-M=\dfrac{\alpha}{2}ln(\dfrac{r}{\mid\alpha\mid})+\dfrac{\alpha}{2}.
\label{E4}
\end{equation}
For $\alpha>0$, the maximum of $\alpha$ satisfies
\begin{equation}
\dfrac{2}{\alpha_{max}}(r-M)-1=ln(\dfrac{2M}{\alpha_{max}}),
\end{equation}
and $\alpha$ is in the range $0<\alpha<2M$.

\noindent For $\alpha<0$, the minimum of $\alpha$ satisfies
\begin{equation}
\dfrac{2}{\alpha_{min}}(r-M)-1=ln(\dfrac{2M}{-\alpha_{min}}),
\end{equation}
and $\alpha$ is in the range $-7.18M<\alpha<0$.

Case II: Positive $\Lambda$ (Kerr-AdS)

There are three horizons, i.e., the inner horizon $r_{-}$, the event horizon $r_{+}$ and the cosmological horizon $r_{\Lambda}$. The horizon equation has two extreme value points. Through the derivative of Eq. (\ref{E2}) with respect to $r$, we get
\begin{equation}
\dfrac{4\Lambda}{3}r^{3}+2(\dfrac{\Lambda}{3}a^{2}-1)r+2M+\alpha=-\alpha ln(\dfrac{r}{\mid\alpha\mid}).
\label{E5}
\end{equation}
For $\alpha>0$, the maximum of $\alpha$ satisfies
\begin{equation}
\alpha_{max}+\alpha_{max} ln(\dfrac{2M}{\alpha_{max}})=2M+H(\Lambda),
\label{E6}
\end{equation}
where $H(\Lambda)=\dfrac{32}{3}\Lambda M^{3}+\dfrac{2}{3}\Lambda a^{2}$, which is a function of $\Lambda$. The existence of the cosmological constant makes  $\alpha_{max}$ to decrease.

\noindent For $\alpha<0$, the minimum of $\alpha$ satisfies
\begin{equation}
\alpha_{min}+\alpha_{min} ln(\dfrac{2M}{-\alpha_{min}})=2M+H(\Lambda).
\label{E6-1}
\end{equation}
The presence of the cosmological constant makes $\alpha_{min}$ to increase.

Case III: Negative $\Lambda$ (Kerr-dS)

There are two horizons, i.e., the inner horizon $r_{-}$ and the event horizon $r_{+}$. The horizon equation has one extreme value point. Through the derivative of Eq.(\ref{E2}) with respect to $r$, we obtain
\begin{equation}
\dfrac{4\Lambda}{3}r^{3}+2(\dfrac{\Lambda}{3}a^{2}+1)r-2M-\alpha=\alpha ln(\dfrac{r}{\mid\alpha\mid}).
\label{E7}
\end{equation}
For $\alpha>0$, the maximum of $\alpha$ satisfies
\begin{equation}
\alpha_{max}+\alpha_{max} ln(\dfrac{2M}{\alpha_{max}})=2M+H(\Lambda),
\label{E8}
\end{equation}
where $H(\Lambda)=-\dfrac{32}{3}\Lambda M^{3}-\dfrac{2}{3}\Lambda a^{2}$. The existence of a negative cosmological constant causes $\alpha_{max}$ to increase.

\noindent For $\alpha<0$, the minimum of $\alpha$ satisfies
\begin{equation}
\alpha_{min}+\alpha_{min} ln(\dfrac{2M}{-\alpha_{min}})=2M+H(\Lambda).
\end{equation}
The existence of a negative cosmological constant causes $\alpha_{min}$ to decrease.

\subsection{Stationary limit surfaces}
\begin{figure}[htbp]
  \centering
  \includegraphics[scale=0.36]{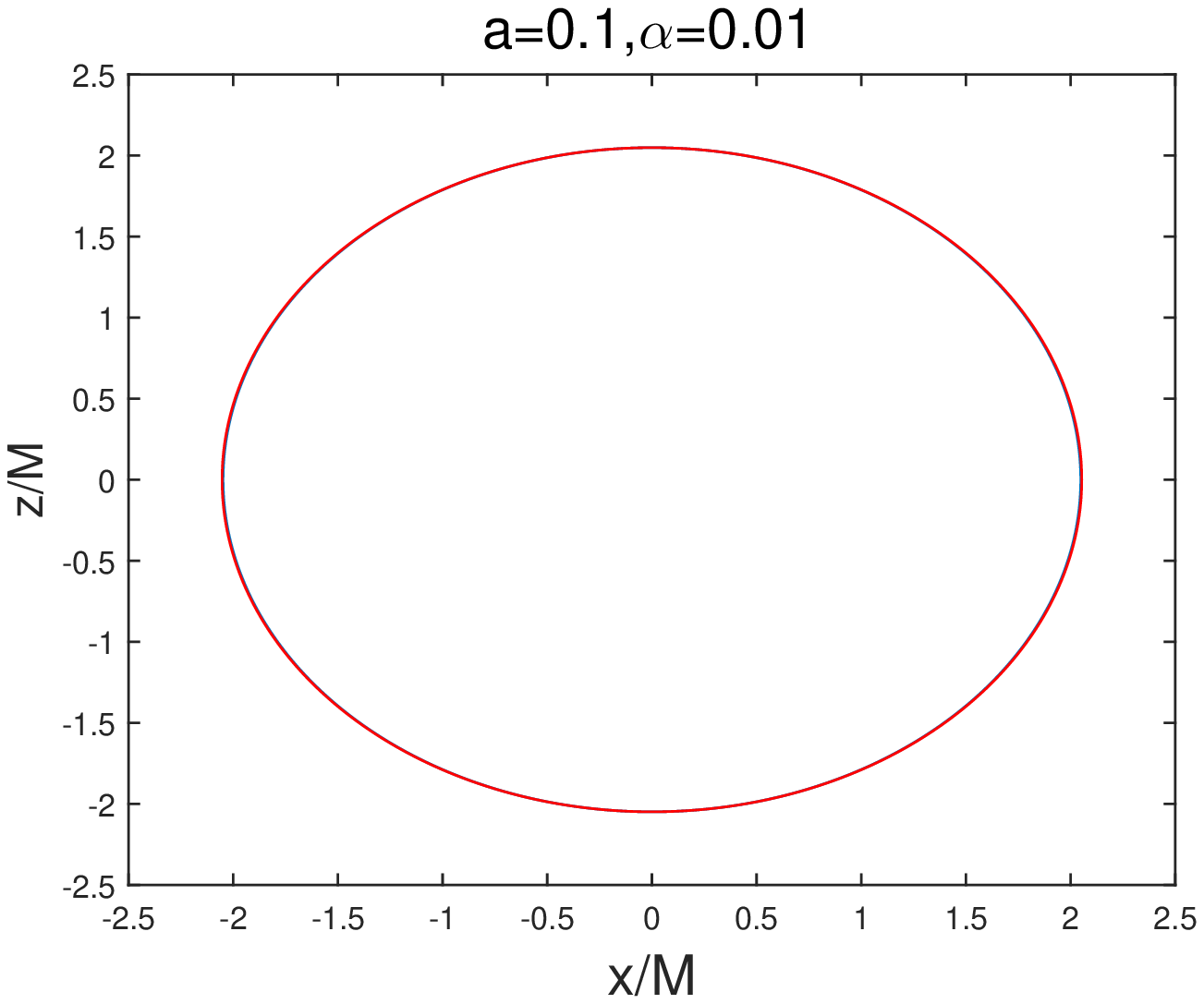}
  \includegraphics[scale=0.36]{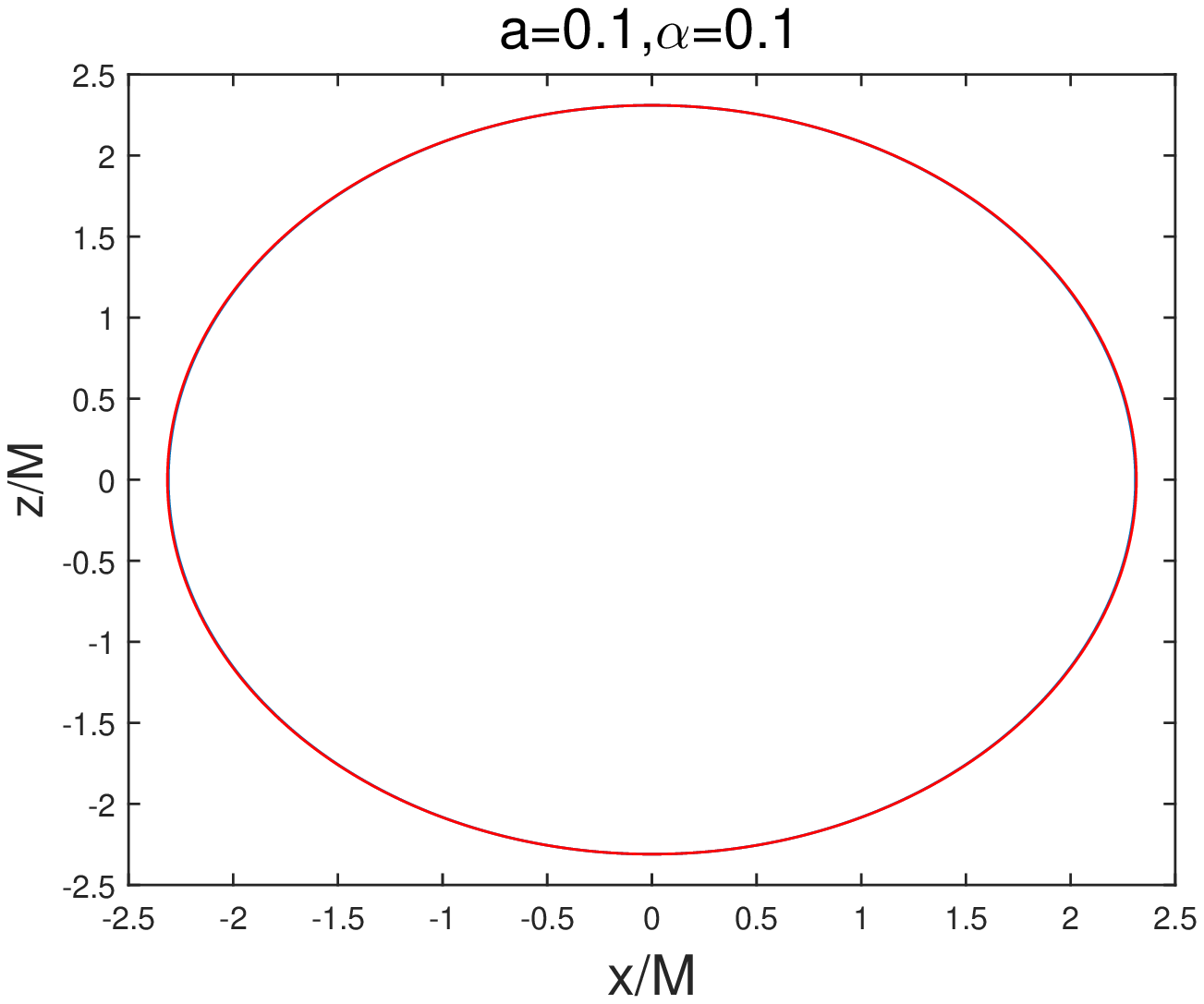}
  \includegraphics[scale=0.36]{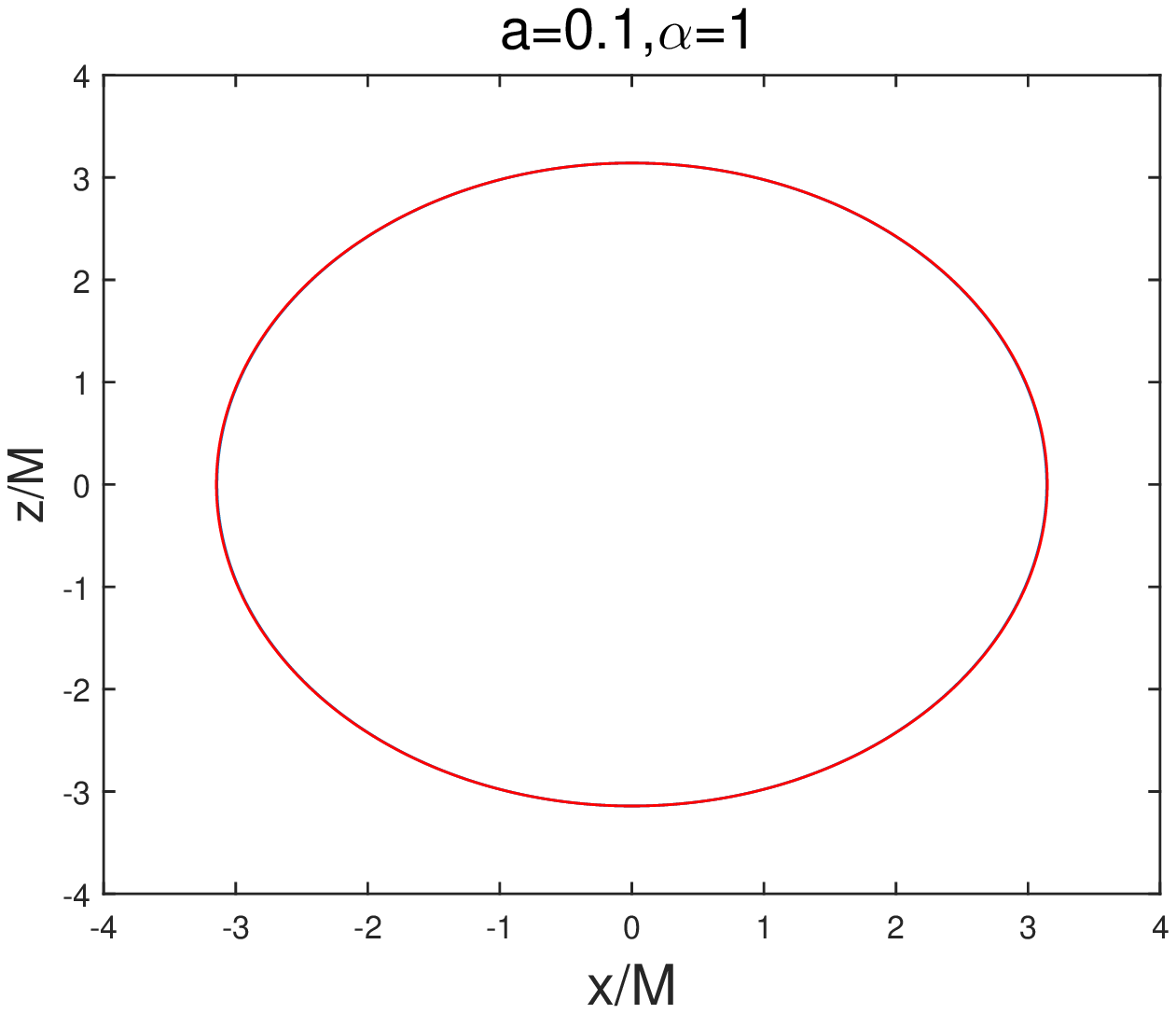}
  \includegraphics[scale=0.36]{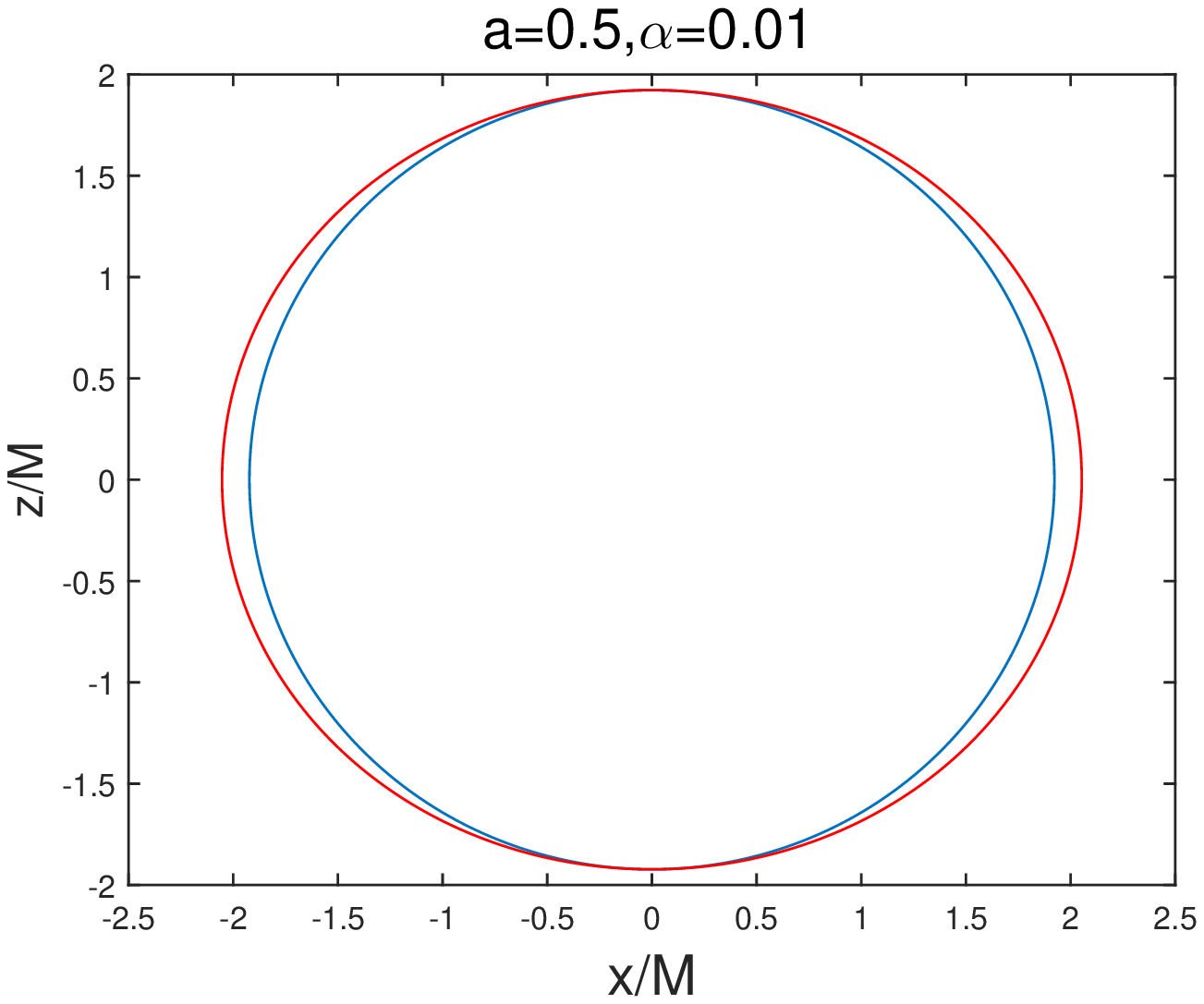}
  \includegraphics[scale=0.36]{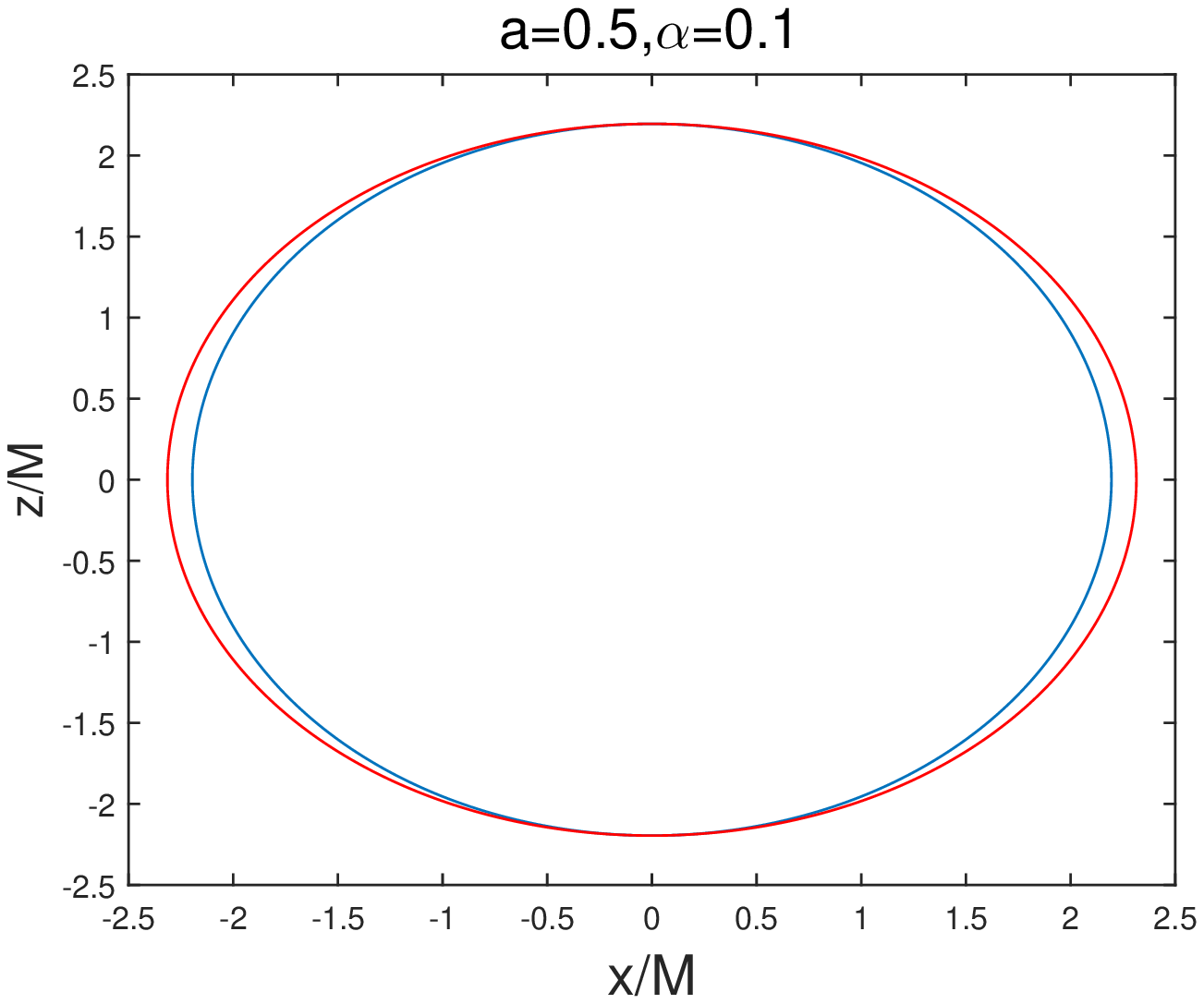}
  \includegraphics[scale=0.36]{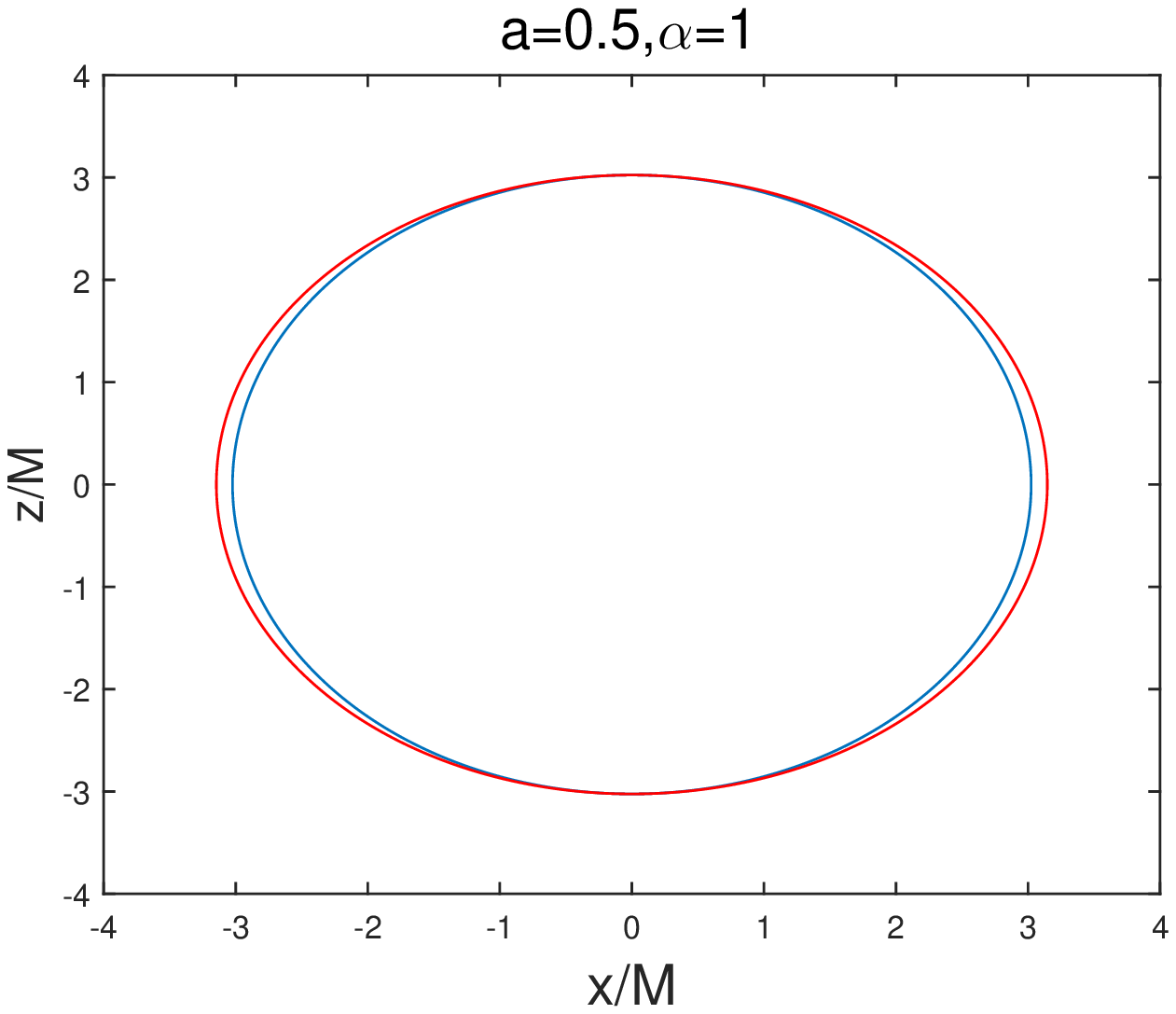}
  \includegraphics[scale=0.36]{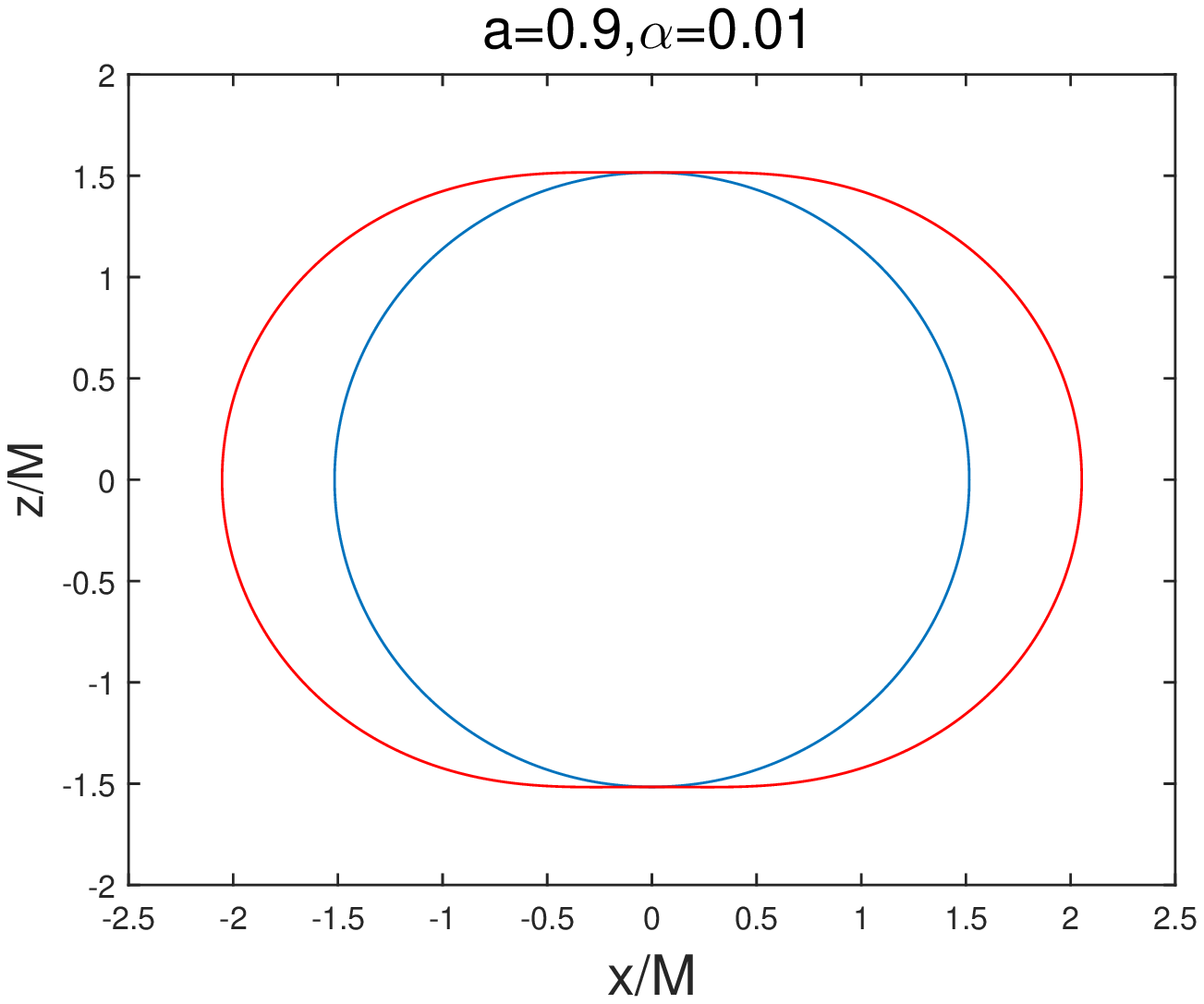}
  \includegraphics[scale=0.36]{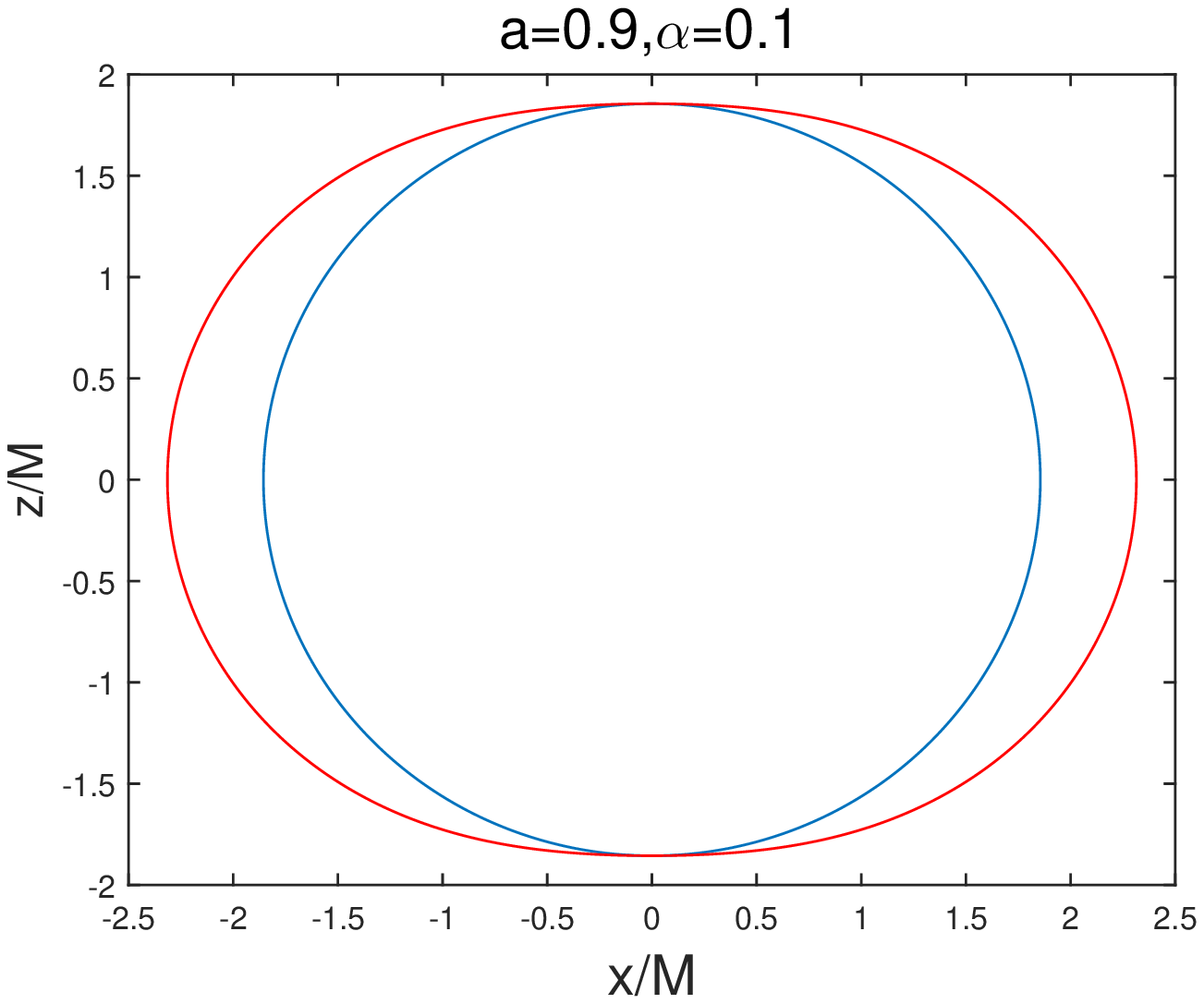}
  \includegraphics[scale=0.36]{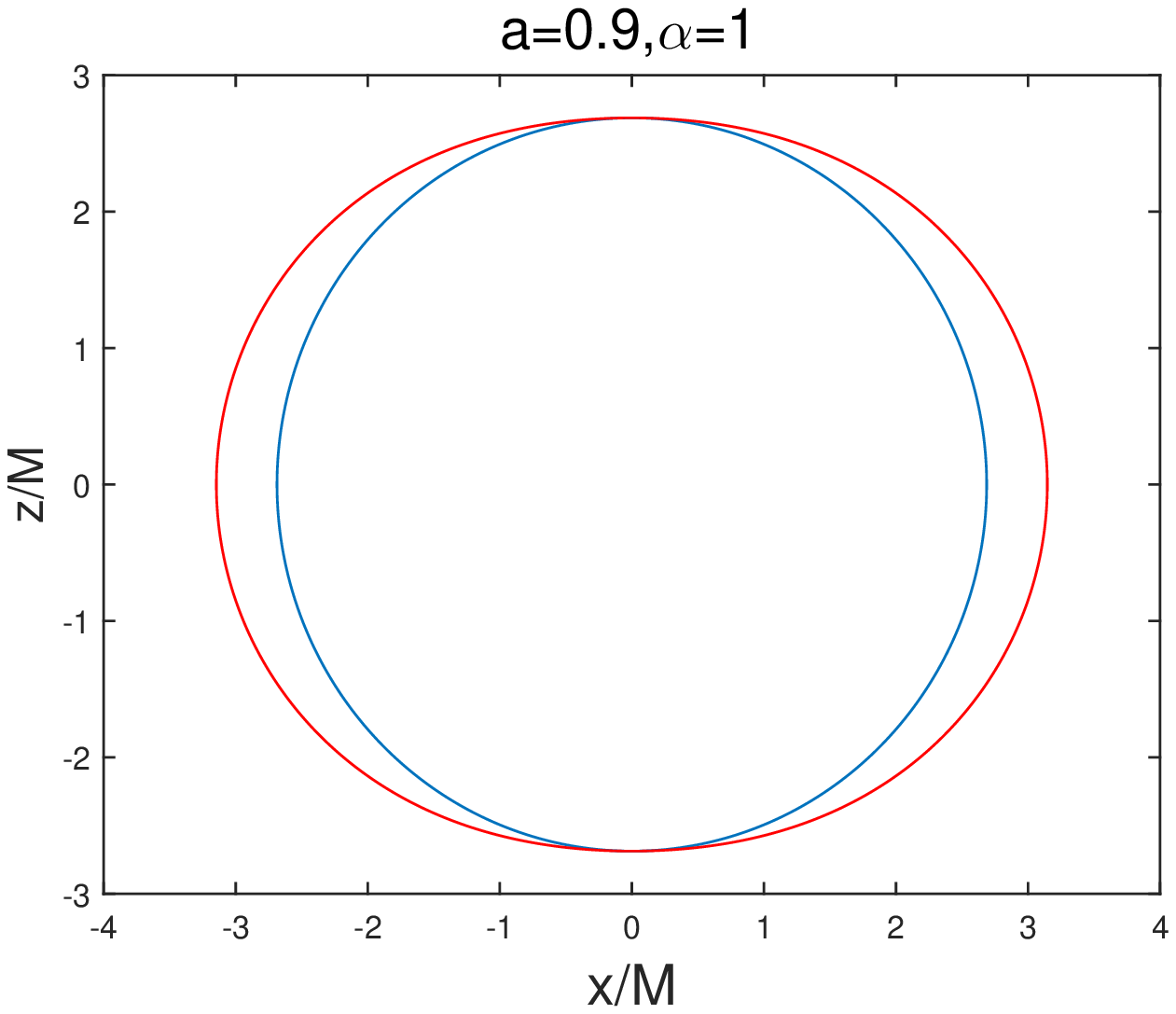}
  \caption{The ergospheres of the Kerr-AdS/dS black hole in perfect fluid DM for different parameters $a$ and $\alpha$ ($\alpha>0$). The blue lines represent the event horizons and the red lines represent the stationary limit surfaces. The region between the event horizon $r_{+}$ and the stationary limit surface $r_{L}$ is the ergosphere. Here we set $\Lambda=\pm1.3\times10^{-56}$ cm$^{-2}\approx 0$. We find that the size of the ergosphere decreases when $\alpha$ increases.}
  \label{fig:1}
\end{figure}

\begin{figure}[htbp]
  \centering
  \includegraphics[scale=0.36]{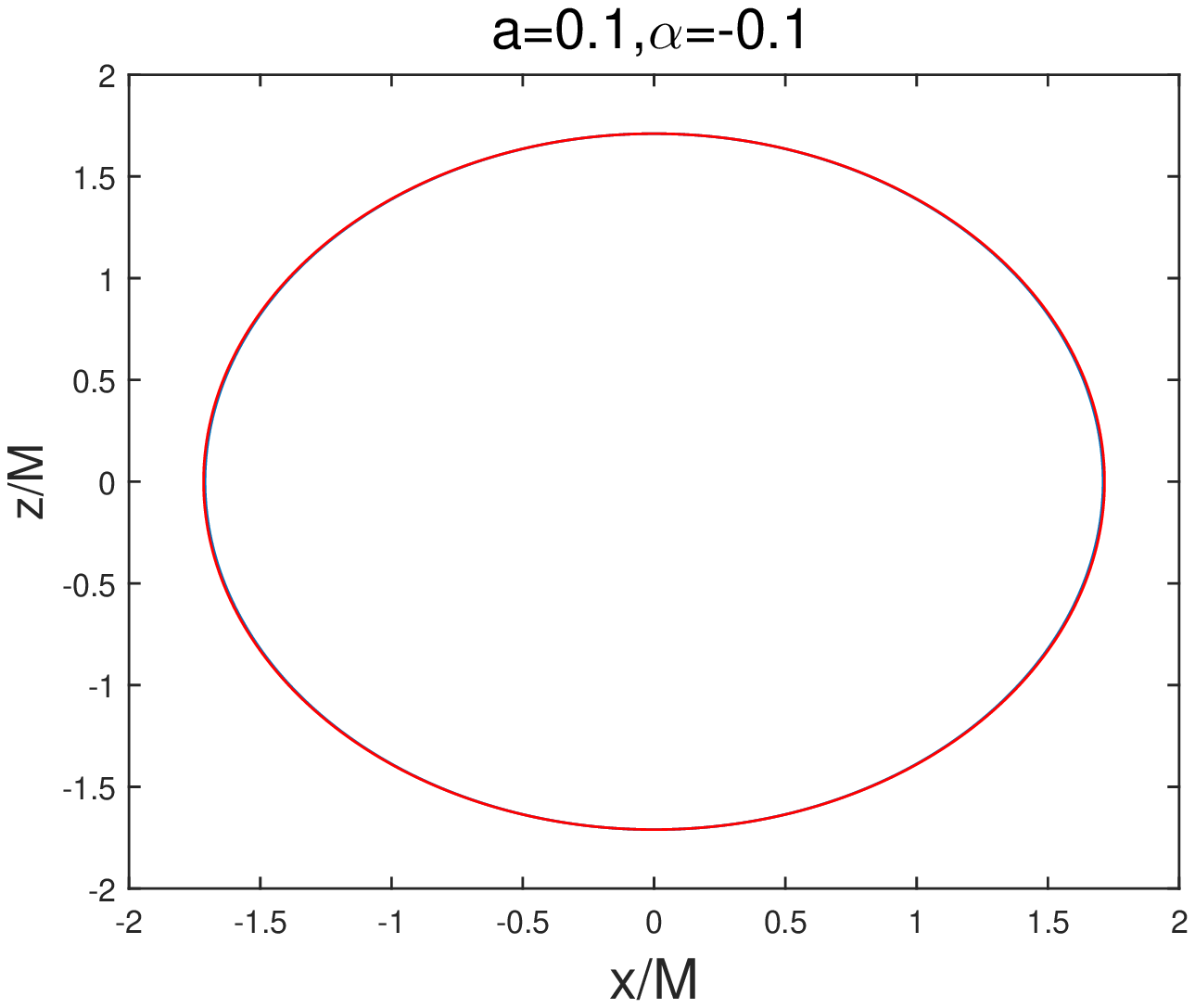}
  \includegraphics[scale=0.36]{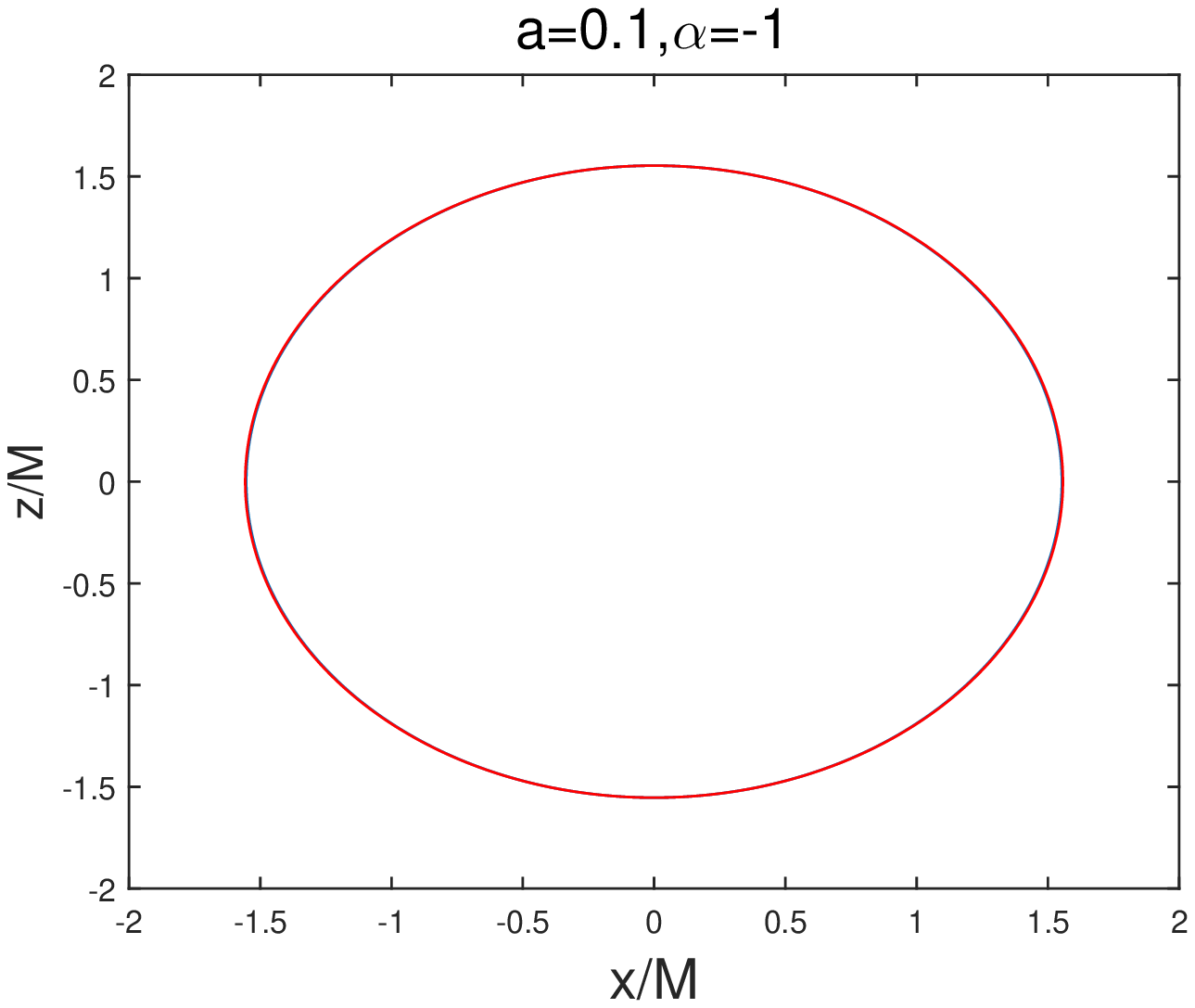}
  \includegraphics[scale=0.36]{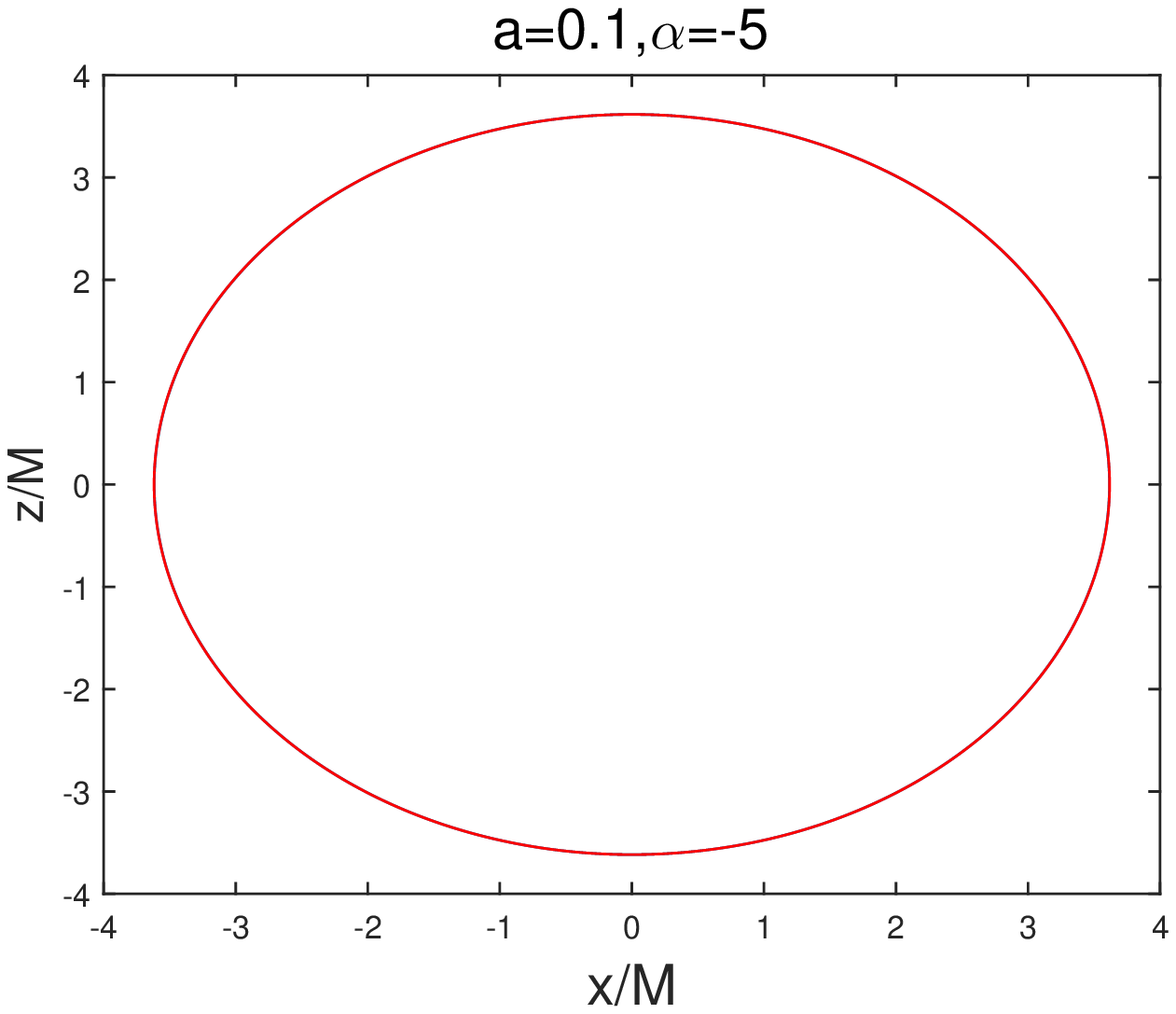}
  \includegraphics[scale=0.36]{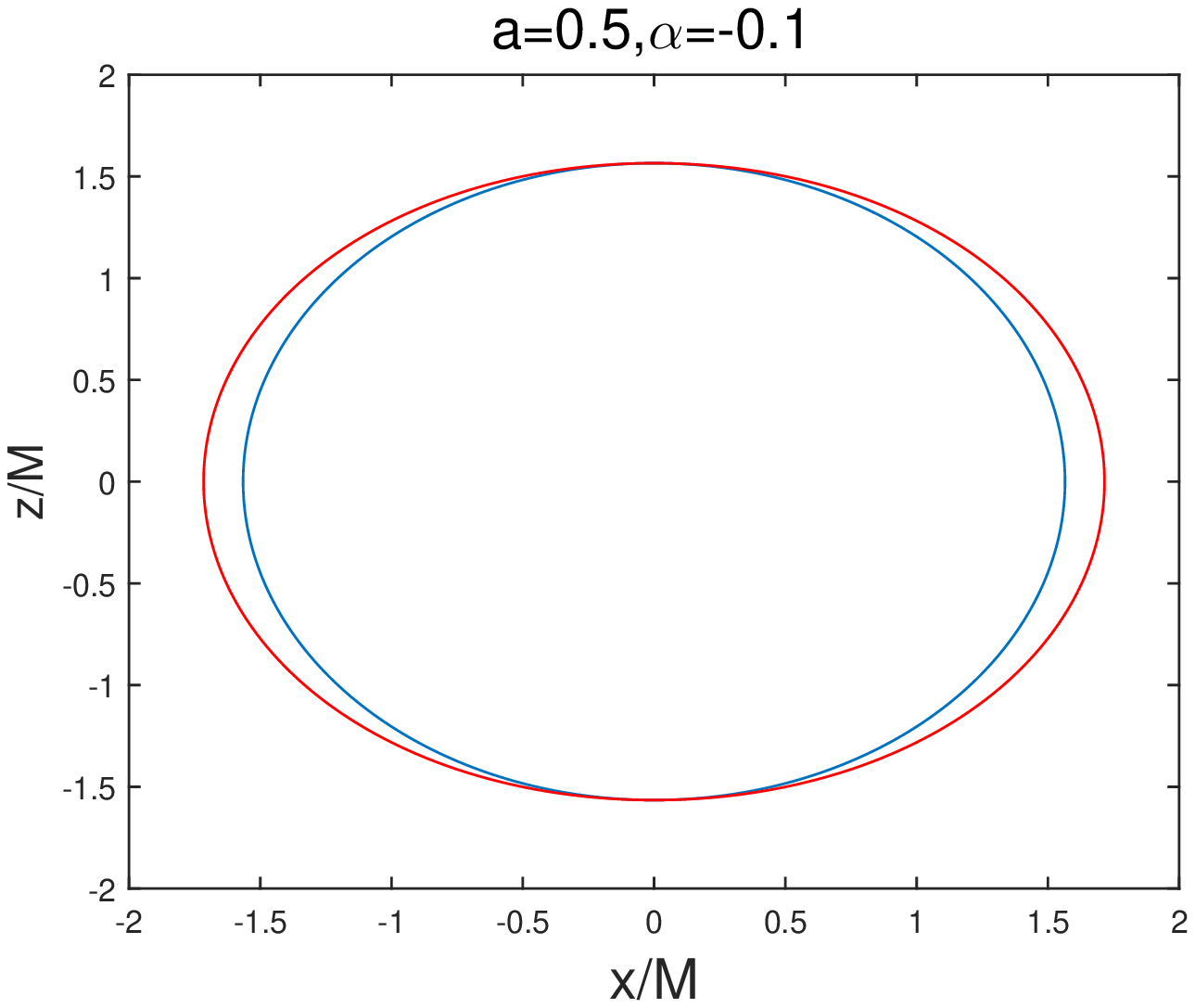}
  \includegraphics[scale=0.36]{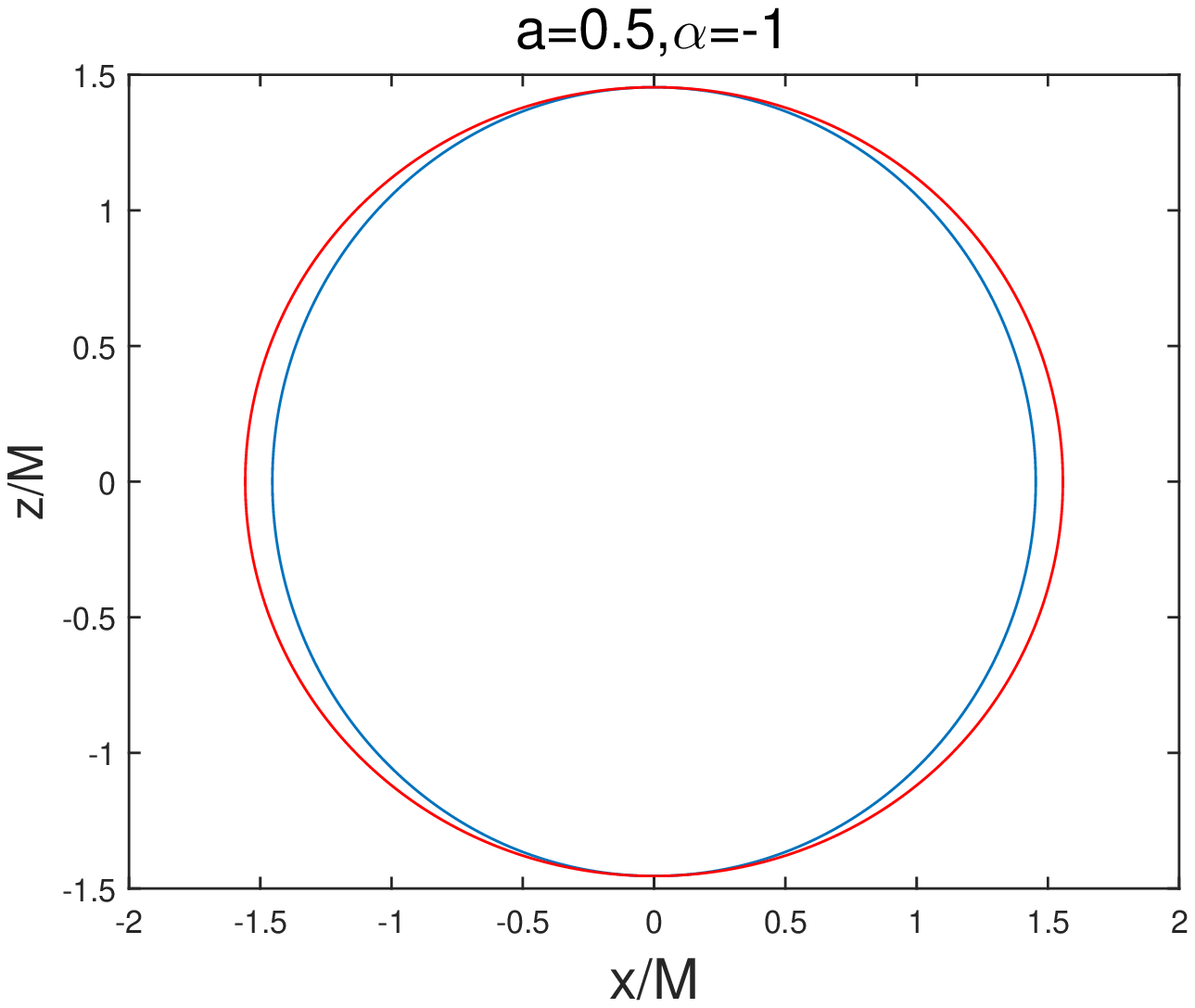}
  \includegraphics[scale=0.36]{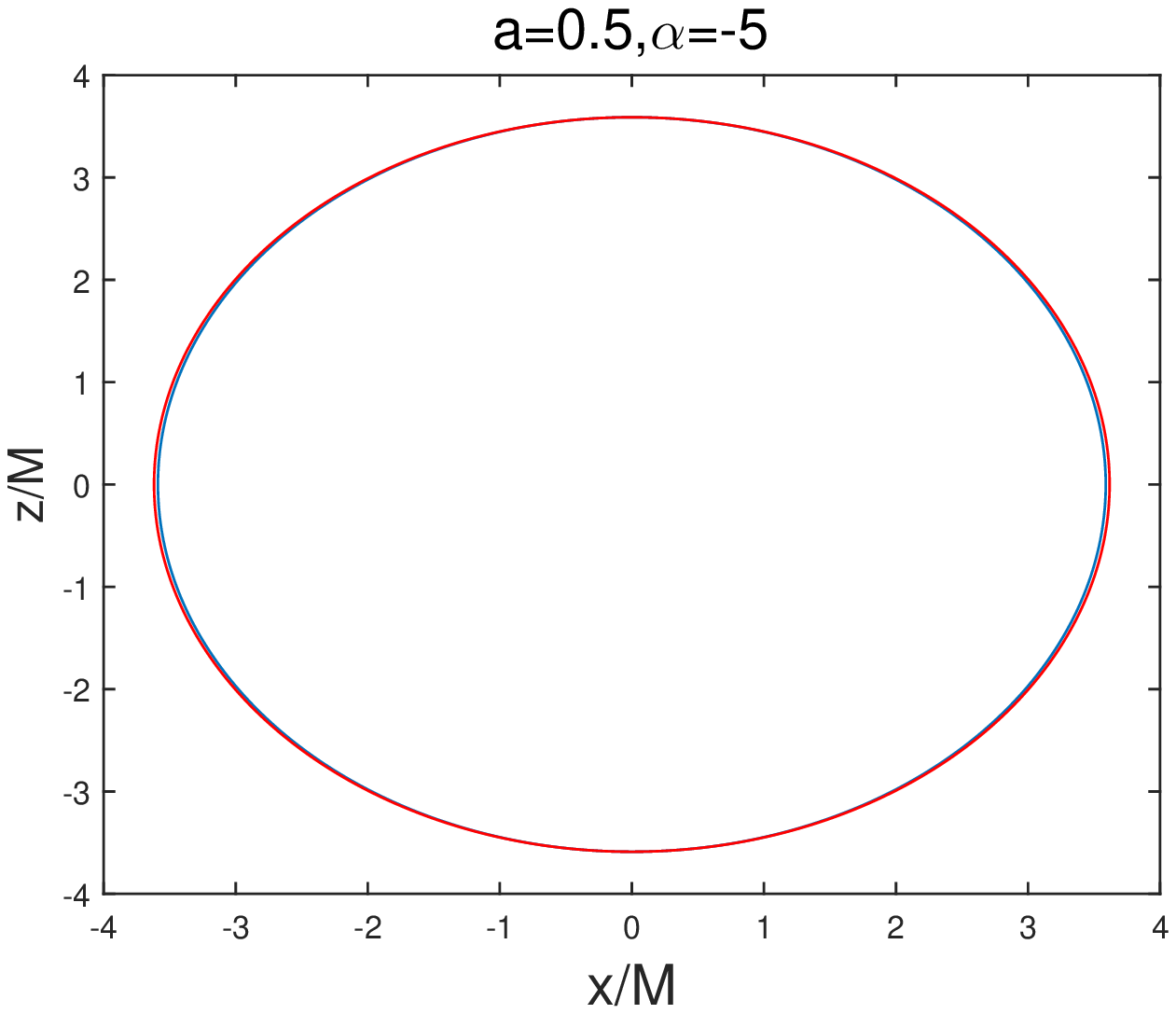}
  \includegraphics[scale=0.36]{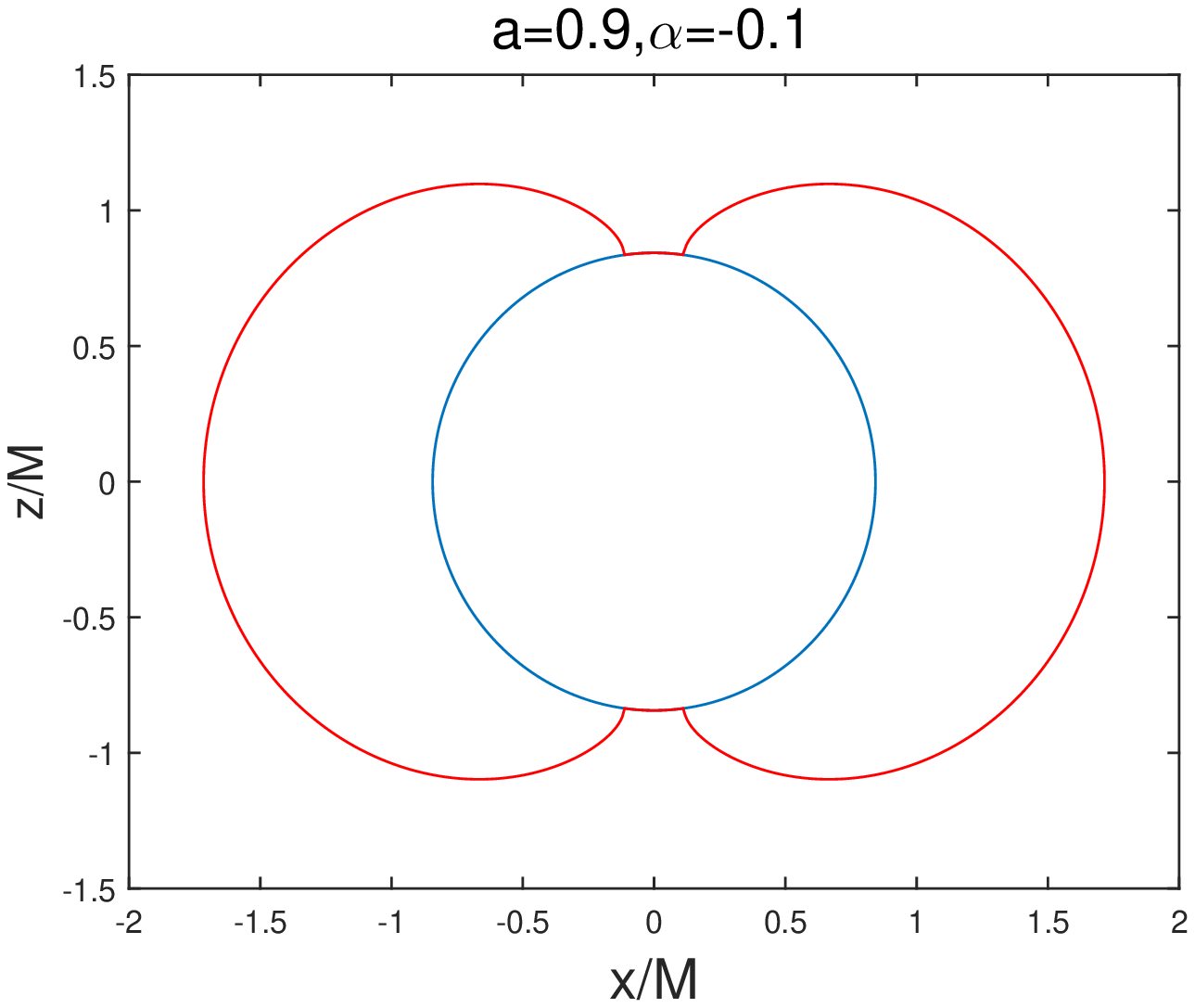}
  \includegraphics[scale=0.36]{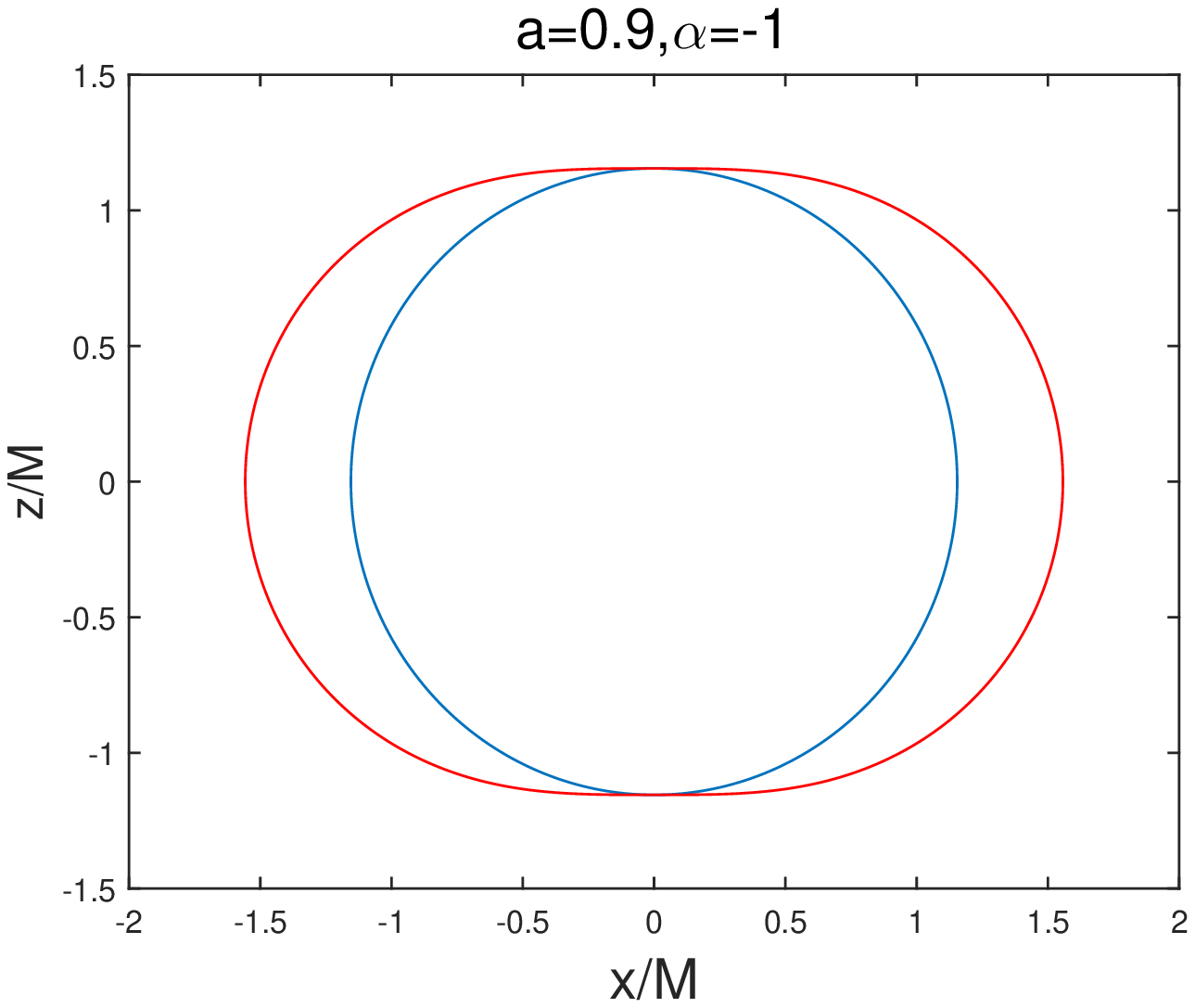}
  \includegraphics[scale=0.36]{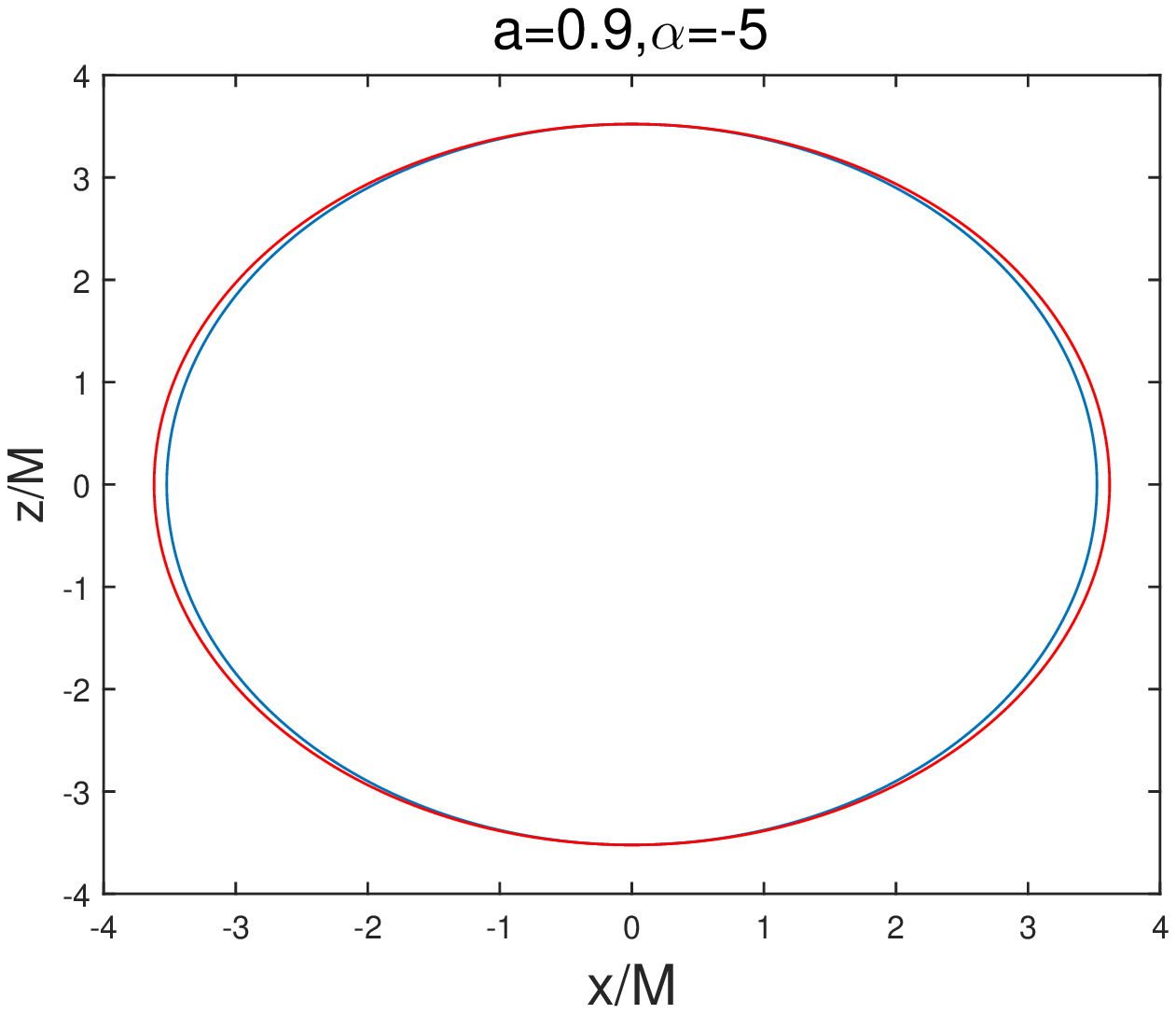}
  \caption{The ergospheres of the Kerr-AdS/dS black hole in perfect fluid DM for different parameters $a$ and $\alpha$ ($\alpha<0$). The blue lines represent the event horizons and the red lines represent the stationary limit surfaces. The region between the event horizon $r_{+}$ and the stationary limit surface $r_{L}$ is the ergosphere. Here we set $\Lambda=\pm1.3\times10^{-56}$ cm$^{-2}\approx 0$. We find that the size of the ergosphere increases when $\alpha$ increases.}
  \label{fig:2}
\end{figure}

The stationary limit surfaces of the rotational black hole are defined by $g_{tt}=0$. From the Einstein field equation (Eq. \ref{AdS3}), the stationary limit surface equation is given by
\begin{equation}
g_{tt}=\dfrac{1}{\Sigma^{2}\Xi^{2}}(a^{2}sin^{2}\theta\Delta_{\theta}-\Delta_{r})$$$$
=\dfrac{1}{\Sigma^{2}\Xi^{2}}(-r^{2}+2Mr+\alpha r ln(\dfrac{r}{\mid\alpha\mid})-a^{2}cos^{2}\theta\pm\dfrac{\Lambda}{3}a^{4}sin^{2}\theta cos^{2}\theta\pm\dfrac{\Lambda}{3}r^{2}(r^{2}+a^{2}))=0.
\label{S1}
\end{equation}
and thus
\begin{equation}
-r^{2}+2Mr+\alpha r ln(\dfrac{r}{\mid\alpha\mid})-a^{2}cos^{2}\theta\pm\dfrac{\Lambda}{3}a^{4}sin^{2}\theta cos^{2}\theta\pm\dfrac{\Lambda}{3}r^{2}(r^{2}+a^{2})=0.
\label{S2}
\end{equation}
By analyzing this equation, we find that there is an ergosphere between the event horizon and the outer static limit surface. The shape of the ergosphere is determined from the parameters $a,\Lambda$ and $\alpha$. Fig.\ref{fig:1} and Fig.\ref{fig:2} show the shapes of the ergospheres for Kerr-AdS/dS black holes surrounded by perfect fluid DM. We find that the increase of $\alpha$ decreases the size of the ergosphere for $\alpha>0$ and increases it for $\alpha<0$. Interestingly, the perfect fluid DM around the black hole has a large effect on the shape of the ergosphere when the dark matter density is high near the black hole. This is in strong contrast to the behaviour of dark energy, which has a rather small effect on the black hole.

\subsection{Singularities}
For the Kerr black hole, we know that there is a singularity ring, and we consider whether or not the singularity changes with the presence of perfect fluid DM. By calculating the Skretshmann scale $R$ defined by $R=R^{\mu\nu\rho\sigma}R_{\mu\nu\rho\sigma}$ from the black hole metric (Eq. \ref{AdS5}), we find that
\begin{equation}
R=R^{\mu\nu\rho\sigma}R_{\mu\nu\rho\sigma}=\dfrac{Z(r,\theta,a,\alpha,\pm\Lambda)}{\Sigma^{12}},
\label{Sing1}
\end{equation}
where $Z(r,\theta,a,\alpha,\pm\Lambda)$ is a polynomial function of $r,\theta,a,\Lambda$ and $\alpha$. The singularity occurs when
\begin{equation}
\Sigma^{2}=r^{2}+a^{2}cos^{2}\theta=0 \Leftrightarrow r=0, \theta=\dfrac{\pi}{2}.
\label{Sing2}
\end{equation}
Because we calculate $R$ in BL coordinates, where $r=0, \theta=\dfrac{\pi}{2}$ represents the ring in the equatorial plane, we find that the result is the same as that for the rotational black hole \citep{1963PhRvL..11..237K} without DM. Therefore the presence of the perfect fluid DM does not change the singularity of black holes.

\section{GEODESIC STRUCTURE IN THE EQUATORIAL PLANE}
\label{geodesic}
We now derive the geodesic motion for the Kerr-AdS/dS black hole surrounded by perfect fluid DM in the equatorial plane. The process is similar to that for the Kerr black hole. We first obtain the separability of variable solutions \citep{1968CMaPh..10..280C} using the Hamiltonian-Jacobi formalism, then we discuss the rotational velocity in the equatorial plane. Note that we do not consider the cosmological constant here because its effect is negligible on the galactic scales that we are interested in. For $\Lambda=0$, we have $\theta=\pi/2$ and $cos\theta=0$. The space-time metric (Eq. \ref{AdS5}) becomes
\begin{equation}
ds^{2}=-(1-\dfrac{2Mr-\alpha r ln(\dfrac{r}{\mid\alpha\mid})}{\Sigma^{2}})dt^{2}+\dfrac{\Sigma^{2}}{\Delta_{r}}dr^{2}-\dfrac{2a sin^{2}\theta(2Mr-\alpha r ln(\dfrac{r}{\mid\alpha\mid}))}{\Sigma^{2}}d\phi dt+\Sigma^{2}d\theta^{2}$$$$
+sin^{2}\theta (r^{2}+a^{2}+a^{2}sin^{2}\theta\dfrac{2Mr-\alpha r ln(\dfrac{r}{\mid\alpha\mid})}{\Sigma^{2}})d\phi^{2},
\label{GS1}
\end{equation}
where
\begin{equation}
\Delta_{r}=r^{2}-2Mr+a^{2}+\alpha r ln(\dfrac{r}{\mid\alpha\mid}).
\label{GS2}
\end{equation}

From the standard method, the Hamilton-Jacobi equation of the Kerr black hole surrounded by perfect fluid DM is given by
\begin{equation}
g^{\mu\nu}\dfrac{\partial S}{\partial x^{\mu}}\dfrac{\partial S}{\partial x^{\nu}}=-m^{2},
\label{HJ1}
\end{equation}
where $m$ is the mass of the test particle. Since the spacetime of black hole surrounded by perfect fluid DM is stationary and axially symmetric, we can define two integrals of the motion, the energy $E=p_{t}$ and the angular momentum $L=-p_{\phi}$, as
\begin{equation}
E=p_{t}=(1-\dfrac{2Mr}{\Sigma^{2}})\dfrac{dt}{d\tau}+\dfrac{2aMr sin^{2}\theta}{\Sigma^{2}}\dfrac{d\phi}{d\tau},
\label{HJ2}
\end{equation}
\begin{equation}
L=-p_{\phi}=-\dfrac{2aMr sin^{2}\theta}{\Sigma^{2}}\dfrac{dt}{d\tau}+(r^{2}+a^{2}+\dfrac{2a^{2}Mr}{\Sigma^{2}}sin^{2}\theta)sin^{2}\theta\dfrac{d\phi}{d\tau}.
\label{HJ3}
\end{equation}
For a test particle following a geodesic of Kerr black hole surrounded by perfect fluid DM, we can set the Hamilton-Jacobi action function as
\begin{equation}
S=-\dfrac{1}{2}m^{2}\tau-Et+L\phi+S_{r}(r)+S_{\theta}(\theta),
\label{HJ4}
\end{equation}
where $\tau$ is the proper time of the test particle, $m^{2}=1$ means the geodesics are timelike, and $m^{2}=0$ corresponds to null geodesics. From the space-time metric of the black hole, we obtain the contravariant components as
\begin{equation}
g^{tt}=-\dfrac{(r^{2}+a^{2})^{2}-a^{2}\Delta_{r}sin^{2}\theta}{\Delta_{r}\Sigma^{2}},~~~~~~g^{rr}=\dfrac{\Delta}{\Sigma^{2}}~~~~g^{\theta\theta}=\dfrac{1}{\Sigma^{2}},$$$$
g^{t\phi}=g^{\phi t}=-\dfrac{2a\rho r}{\Delta_{r}\Sigma^{2}},~~~~g^{\phi\phi}=\dfrac{\Delta_{r}-a^{2}sin^{2}\theta}{\Delta_{r}\Sigma^{2}sin^{2}\theta}.
\label{HJ5}
\end{equation}
Combining Eq. (\ref{HJ1}), (\ref{HJ4}) and (50), we obtain
\begin{equation}
-m^{2}\Sigma^{2}=\dfrac{1}{\Delta_{r}}((r^{2}+a^{2})E-aL)^{2}-\dfrac{1}{sin^{2}\theta}(aE sin^{2}\theta-L)^{2}-\Delta_{r}(\dfrac{dS_{r}}{dr})^{2}-(\dfrac{dS_{\theta}}{d\theta})^{2}.
\label{HJ6}
\end{equation}
Considering the identity of
\begin{equation}
(aE sin^{2}\theta-L)^{2}cosec^{2}\theta=(L^{2}cosec^{2}\theta-a^{2}E^{2})cos^{2}\theta+(L-aE)^{2},
\label{HJ7}
\end{equation}
we find that Eq. (\ref{HJ6}) becomes
\begin{equation}
[\Delta_{r}(\dfrac{dS_{r}}{dr})^{2}-\dfrac{1}{\Delta_{r}}((r^{2}+a^{2})E-aL)^{2}+(L-aE)^{2}-m^{2}r^{2}]+$$$$
[(\dfrac{dS_{\theta}}{d\theta})^{2}+(L^{2}cosec^{2}\theta-a^{2}E^{2})cos^{2}\theta-m^{2}a^{2}cos^{2}\theta]=0.
\label{HJ8}
\end{equation}
By simplifying Eq. (53), introducing the separation constant $K$ and defining a new constant of the motion through the relation $\Pi=K-(L-aE)$, we get
\begin{equation}
\Delta_{r}(\dfrac{dS_{r}}{dr})^{2}=\dfrac{1}{\Delta_{r}}((r^{2}+a^{2})E-aL)^{2}-(\Pi+(L-aE)^{2}-m^{2}r^{2}),
\label{HJ9}
\end{equation}
and
\begin{equation}
(\dfrac{dS_{\theta}}{d\theta})^{2}=\Pi-(L^{2}cosec^{2}\theta-a^{2}E^{2}-m^{2}a^{2})cos^{2}\theta.
\label{HJ10}
\end{equation}
Setting
\begin{equation}
R(r)=((r^{2}+a^{2})E-aL)^{2}-\Delta_{r}(\Pi+(L-aE)^{2}-m^{2}r^{2}),
\label{HJ11}
\end{equation}
and
\begin{equation}
\Theta(\theta)=\Pi-(L^{2}cosec^{2}\theta-a^{2}E^{2}-m^{2}a^{2})cos^{2}\theta,
\label{HJ12}
\end{equation}
leads to
\begin{equation}
S=-\dfrac{1}{2}m^{2}\tau-Et+L\phi+\int\dfrac{\sqrt{R(r)}}{\Delta_{r}}dr+\int\sqrt{\Theta(\theta)}dr.
\label{HJ13}
\end{equation}
Following the original paper \citep{1968CMaPh..10..280C}, we obtain the geodesic equation of motion as
\begin{equation}
\Sigma^{2}\dfrac{dt}{d\tau}=\dfrac{r^{2}+a^{2}}{\Delta_{r}}((r^{2}+a^{2})E-aL)-a(aE sin^{2}\theta-L),
\label{HJ14}
\end{equation}
\begin{equation}
\Sigma^{2}\dfrac{dr}{d\tau}=\sqrt{R(r)},
\label{HJ15}
\end{equation}
\begin{equation}
\Sigma^{2}\dfrac{d\theta}{d\tau}=\sqrt{\Theta(\theta)},
\label{HJ16}
\end{equation}
\begin{equation}
\Sigma^{2}\dfrac{d\phi}{d\tau}=\dfrac{a}{\Delta_{r}}((r^{2}+a^{2})E-aL)-(aE-\dfrac{L}{sin^{2}\theta}).
\label{HJ17}
\end{equation}

If we consider these solutions in the equatorial plane, where $\theta=\pi/2$, $d\theta/d\tau=0$, $\Theta(\pi/2)=0$ and $\Pi=0$, the radial motion of the test particle is given by
\begin{equation}
r^{2}\dfrac{dr}{d\tau}=\pm\sqrt{R(r)}=\pm\sqrt{((r^{2}+a^{2})E-aL)^{2}-\Delta_{r}((L-aE)^{2}-m^{2}r^{2})}.
\label{RV1}
\end{equation}
The stable circular orbit must satisfy the two conditions
\begin{equation}
R(r)=0,~~~~~~~~\dfrac{dR(r)}{dr}=0,
\label{RV2}
\end{equation}
and we get
\begin{equation}
((r^{2}+a^{2})E-aL)^{2}-\Delta_{r}((L-aE)^{2}-m^{2}r^{2})=0,
\label{RV3}
\end{equation}
 \begin{equation}
4rE((r^{2}+a^{2})E-aL)-\dfrac{d\Delta_{r}}{dr}((L-aE)^{2}+m^{2}r^{2})-2m^{2}r\Delta=0.
\label{RV4}
\end{equation}
Finally we obtain expressions for the energy and angular momentum, respectively, as
\begin{equation}
E^{2}_{\pm}=m^{2}\dfrac{f_{1}(r,a,\alpha,\Delta_{r})\pm \sqrt{(2a^{2}-2\Delta_{r}+r\dfrac{d\Delta_{r}}{dr})^{3}}g_{1}(r,a,\alpha,\Delta_{r})}{r^{2}(16\Delta_{r}(\Delta_{r}-a^{2})+r\dfrac{d\Delta_{r}}{dr}(r\dfrac{d\Delta_{r}}{dr}-8\Delta_{r}))},
\label{RV5}
\end{equation}
\begin{equation}
L^{2}_{\pm}=m^{2}\dfrac{f_{2}(r,a,\alpha,\Delta_{r})\mp \sqrt{(2a^{2}-2\Delta_{r}+r\dfrac{d\Delta_{r}}{dr})^{3}}g_{2}(r,a,\alpha,\Delta_{r})}{r^{2}(16\Delta_{r}(\Delta_{r}-a^{2})+r\dfrac{d\Delta_{r}}{dr}(r\dfrac{d\Delta_{r}}{dr}-8\Delta_{r}))},
\label{RV6}
\end{equation}
where $f_{1}(r,a,\alpha,\Delta_{r}), f_{2}(r,a,\alpha,\Delta_{r}), g_{1}(r,a,\alpha,\Delta_{r})$ and $g_{2}(r,a,\alpha,\Delta_{r})$ are functions of $r,a,\alpha$ and $\Delta_{r}$. If the test particle moves along a circular orbit, $E$ and $L$ must be real, and the following condition
\begin{equation}
2a^{2}-2\Delta_{r}+r\dfrac{d\Delta_{r}}{dr}=2Mr+\alpha r ln(\dfrac{r}{\mid\alpha\mid})-\alpha r\geq 0
\label{RV7}
\end{equation}
should be satisfied. For Kerr black holes surrounded by perfect fluid DM, we have $\Lambda=0$ and $r=2M$. If $\alpha>0$, Eq. (\ref{RV7}) becomes $\dfrac{2}{\alpha}(r-M)-1-ln(\dfrac{2M}{\alpha})\geq 0$; If $\alpha<0$, Eq. (\ref{RV7}) becomes $\dfrac{2}{\alpha}(r-M)-1-ln(\dfrac{2M}{-\alpha})\geq 0$. These results are same as those presented in Section \ref{properties}.

The rotational velocity of the test particle in the equatorial plane is given by
\begin{equation}
v=\dfrac{L}{\sqrt{g_{\phi\phi}}}.
\label{RV20}
\end{equation}

\begin{figure}[htbp]
  \centering
  \includegraphics[scale=0.47]{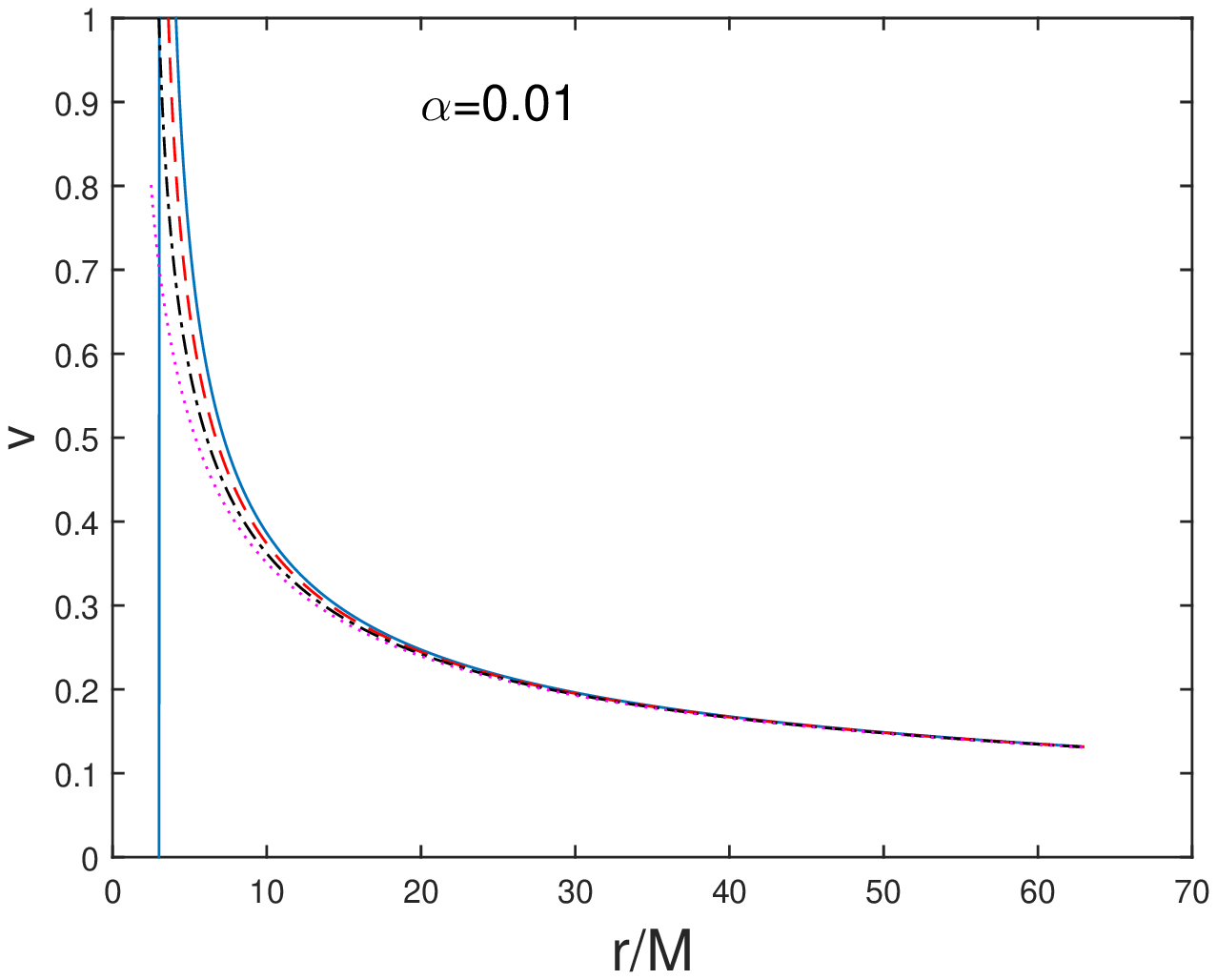}
  \includegraphics[scale=0.47]{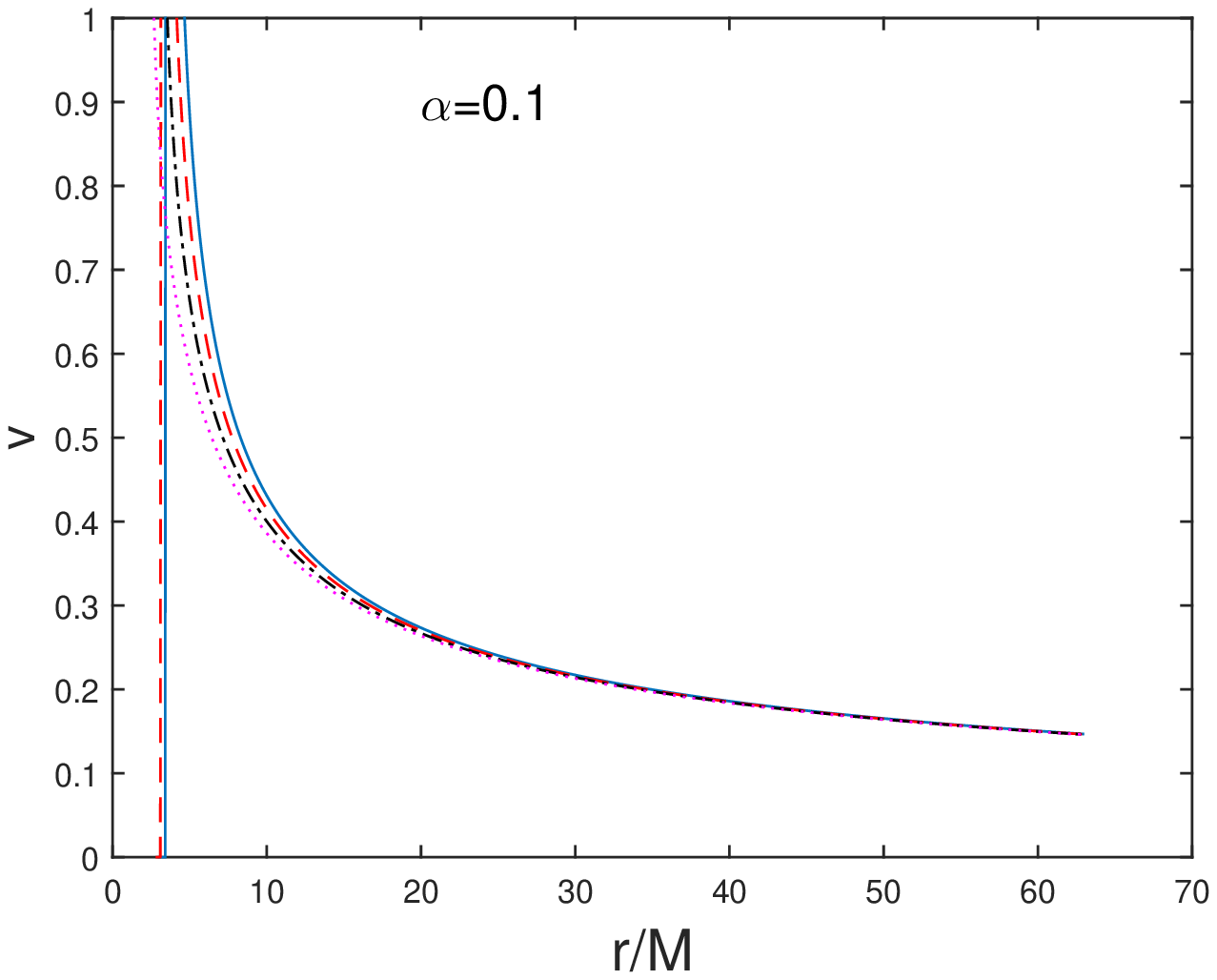}
  \includegraphics[scale=0.47]{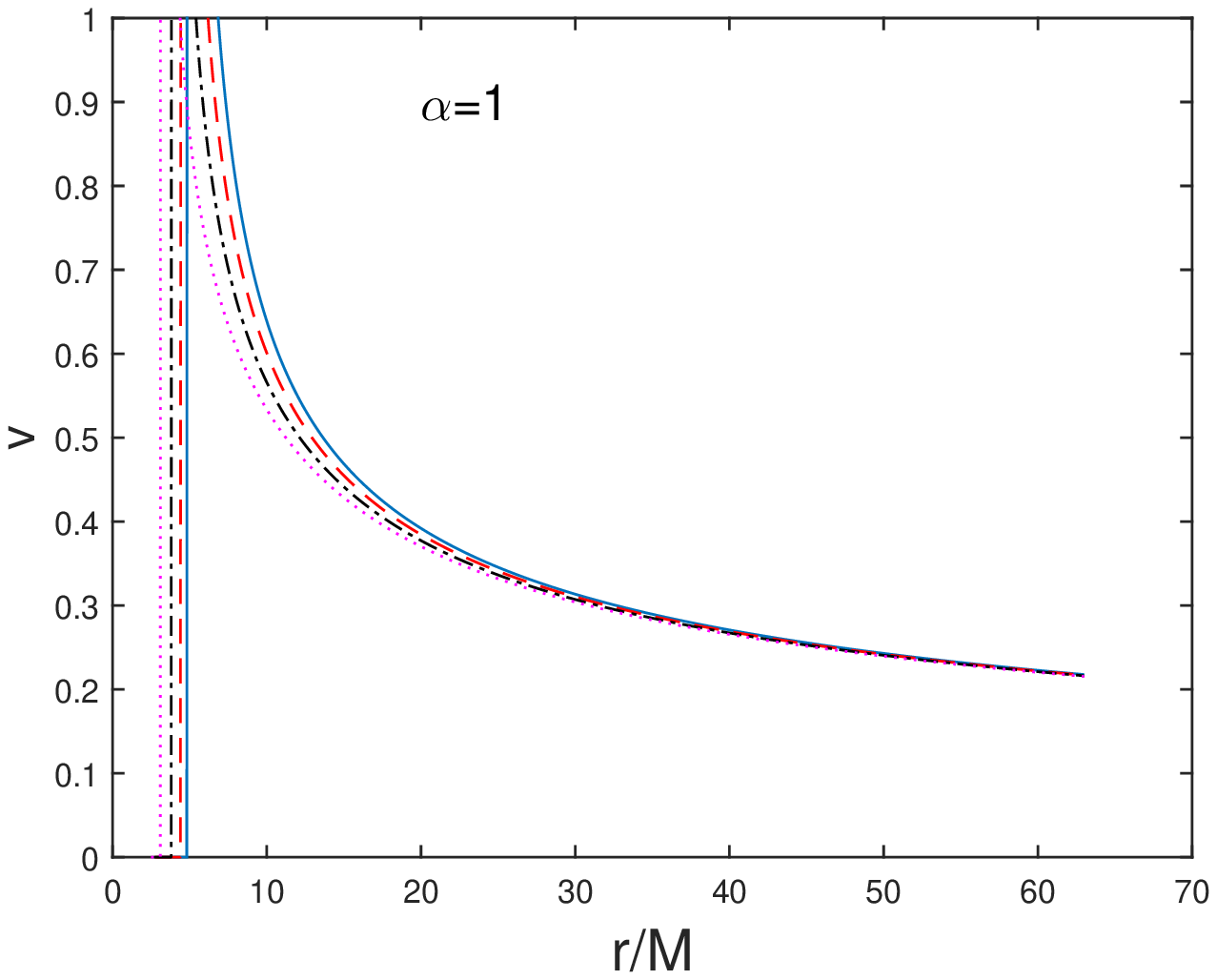}
  \includegraphics[scale=0.47]{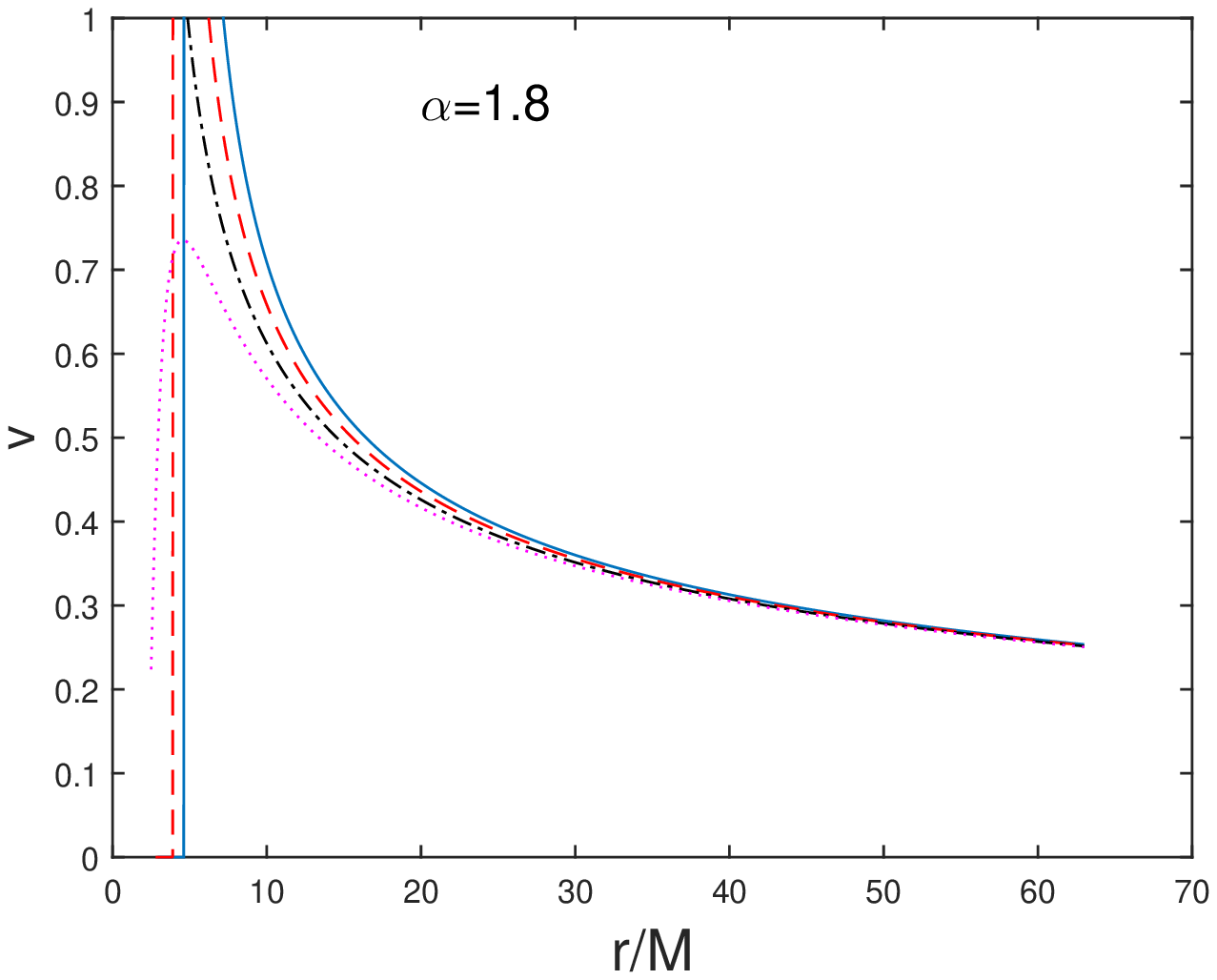}
  \caption{The behaviour of the rotational velocity $v$ as a function of  $r$ in the equatorial plane of the Kerr black hole with the presence of perfect fluid DM ($\alpha>0$) for different parameters $a$ and $\alpha$. Solid line: $a=0$, dashed line: $a=0.3$, dotdashed line: $a=0.6$, dotted line: $a=0.9$.}
  \label{fig:3}
\end{figure}

\begin{figure}[htbp]
  \centering
  \includegraphics[scale=0.47]{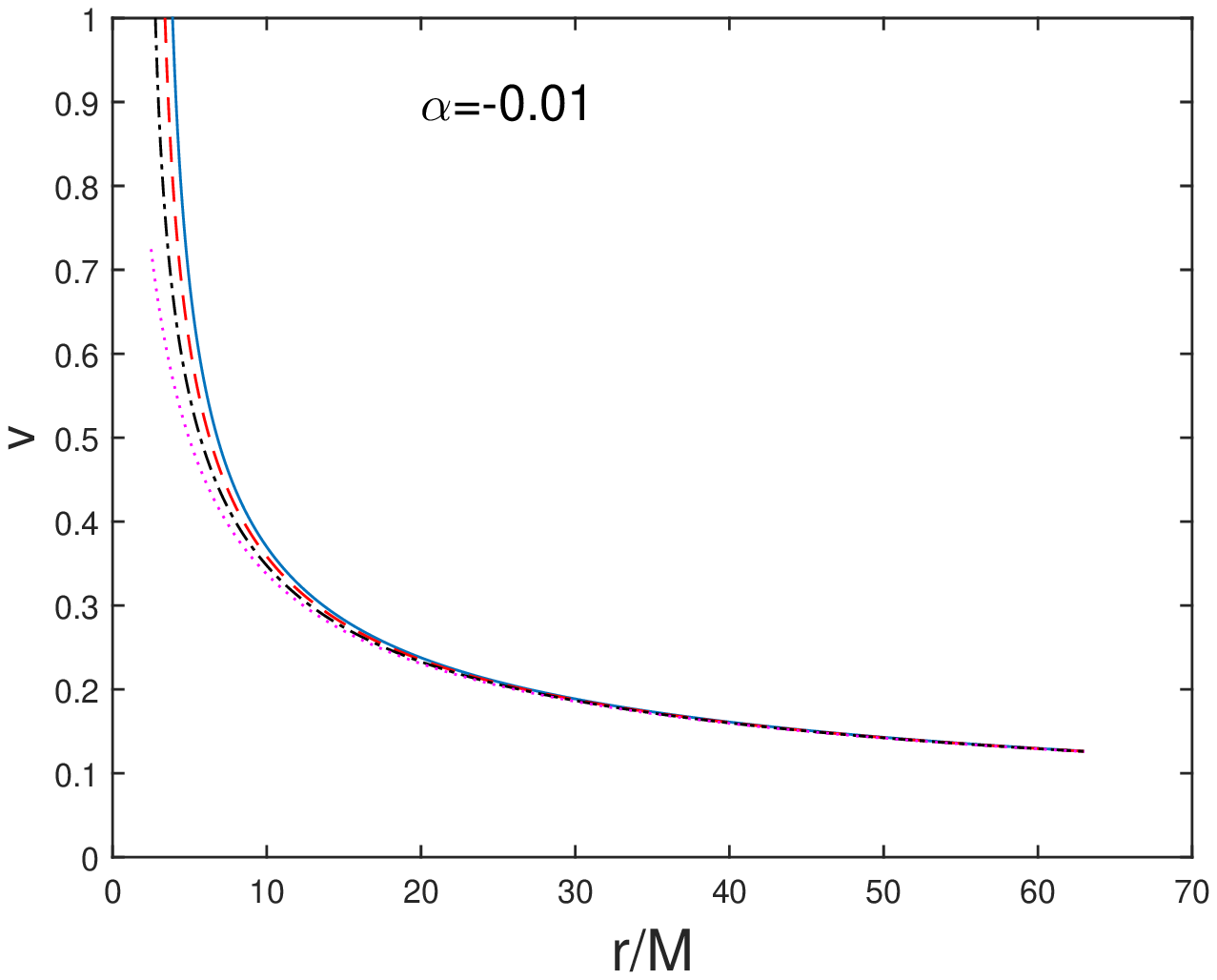}
  \includegraphics[scale=0.47]{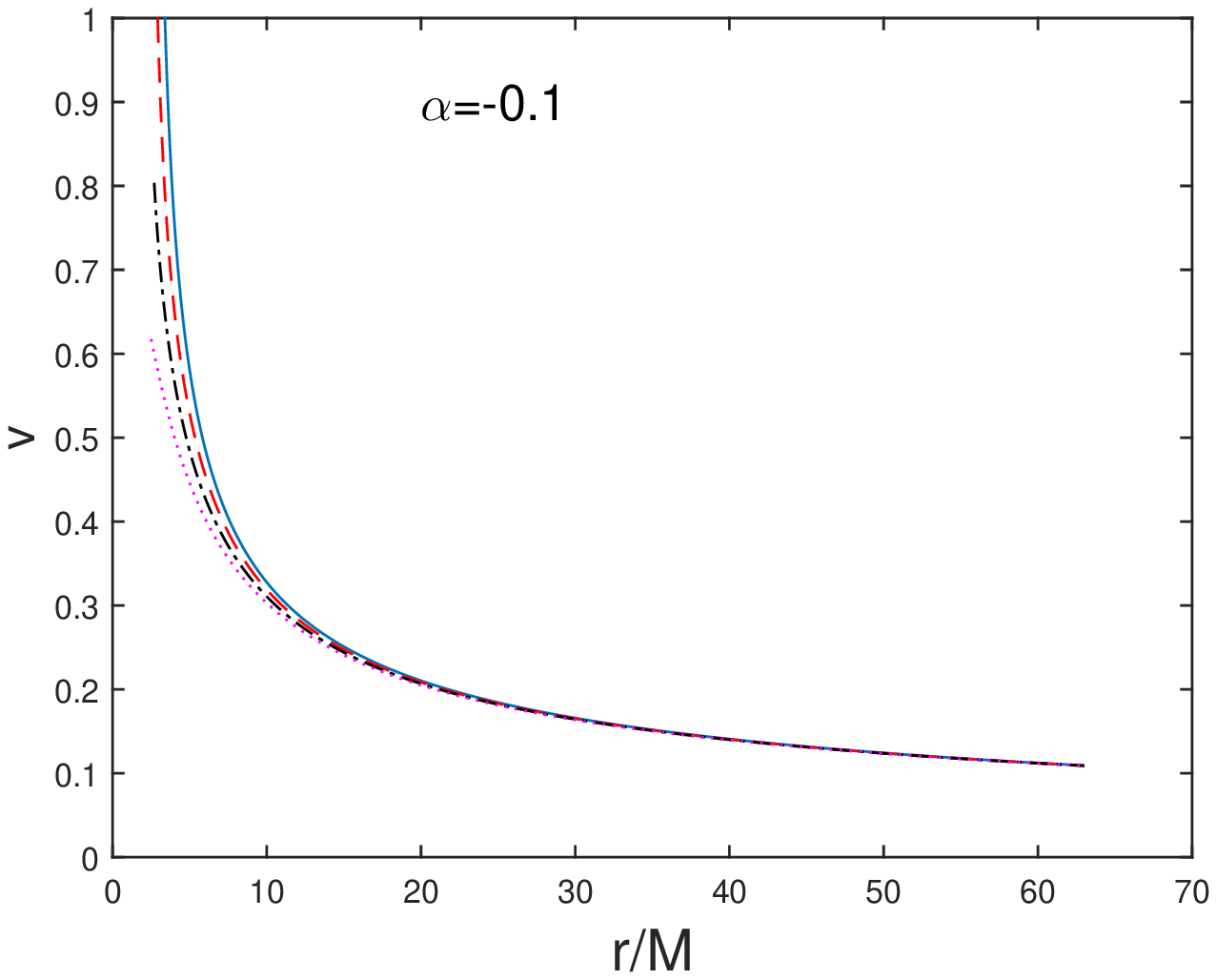}
  \includegraphics[scale=0.47]{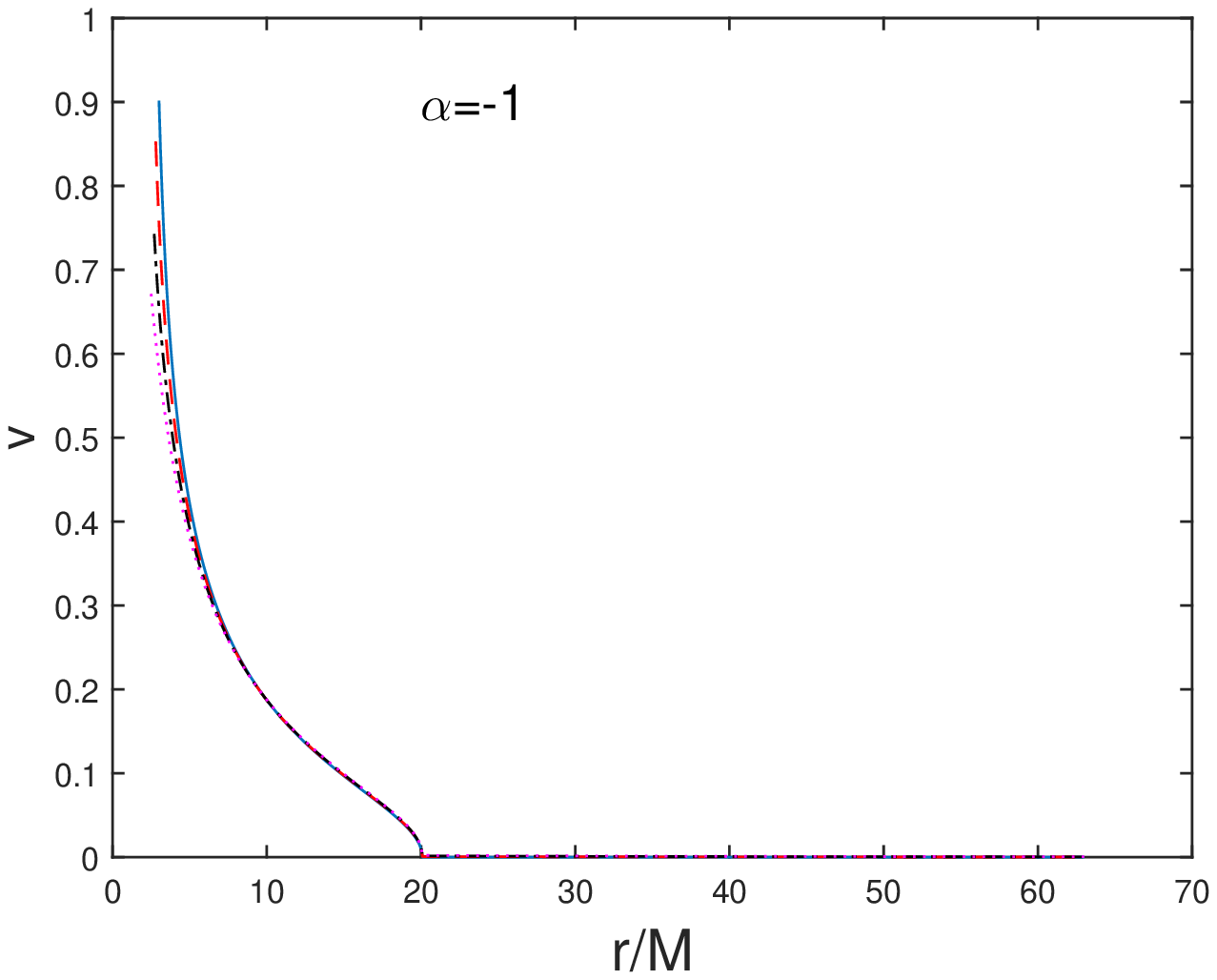}
  \includegraphics[scale=0.47]{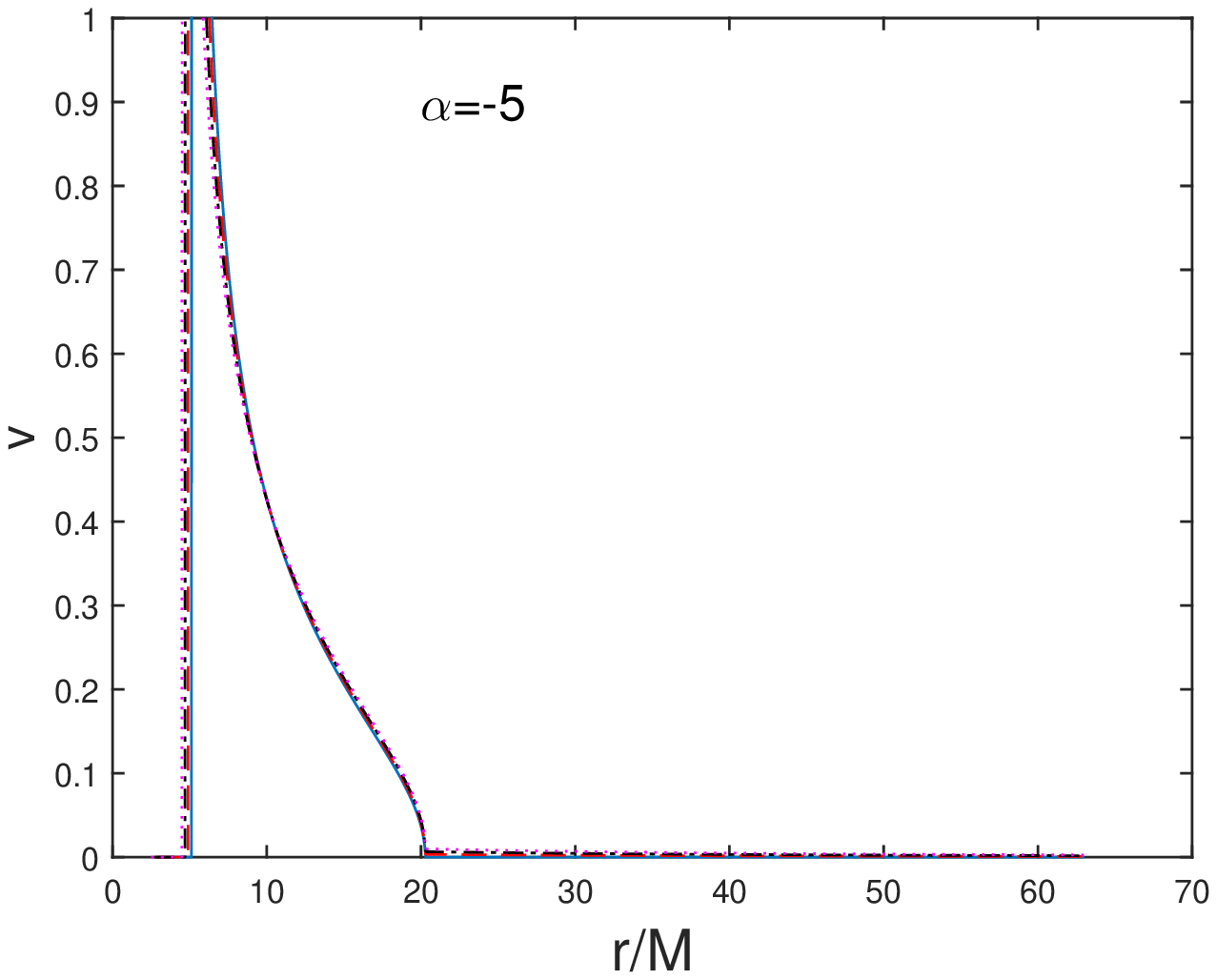}
  \caption{The behaviour of the rotational velocity $v$ as a function of  $r$ in the equatorial plane of the Kerr black hole with the presence of perfect fluid DM ($\alpha<0$) for different parameters $a$ and $\alpha$. Solid line: $a=0$, dashed line: $a=0.3$, dotdashed line: $a=0.6$, dotted line: $a=0.9$.}
  \label{fig:4}
\end{figure}

Fig.\ref{fig:3} and Fig.\ref{fig:4} show the behaviour of the rotational velocity $v$ with respect to $r$ in the equatorial plane of the Kerr black hole in perfect fluid DM. We find that: (1) if $\alpha>0$, then the rotational velocity is asymptotically flat and independent of $\alpha$; (2) if $\alpha<0$, then the rotational velocity is asymptotically flat only when $\alpha$ is close to zero.

\section{SUMMARY}
\label{summary}
We obtain the solution of the Kerr black hole surrounded by the perfect fluid dark matter using the Newman-Janis method and generalize it to include a cosmological constant. By analyzing the horizon equation and the static circular orbit, we obtain the relation between the dark matter parameter $\alpha$ and the positive cosmological constant when the cosmological horizon $r_{\Lambda}$ exists. For $\Lambda=0$, we find that $\alpha$ is in the range of $0<\alpha<2M$ for $\alpha>0$ and $-7.18M<\alpha<0$ for $\alpha<0$. For positive cosmological constant $\Lambda$, $\alpha_{max}$ decreases for $\alpha>0$ while $\alpha_{min}$ increases for $\alpha<0$. For negative cosmological constant $-\Lambda$, $\alpha_{max}$ increases for $\alpha>0$ and $\alpha_{min}$ decreases for $\alpha<0$. The size of the ergosphere evolves oppositely for $\alpha>0$ and $\alpha<0$, while the ergosphere size decreases with increasing $\mid\alpha\mid$. The singularity is the same as that of the Kerr black hole. We also study the geodesic motion using the Hamilton-Jacobi formalism and analyze the rotational velocity of the black hole in the equatorial plane. If $\alpha$ is in the above ranges for $\Lambda=0$, stable orbits exist. The rotational velocity is asymptotically flat and independent of $\alpha$ if $\alpha>0$ while is asymptotically flat only when $\alpha$ is close to zero if $\alpha<0$.

Kerr-AdS/dS black holes surrounded by perfect fluid dark matter could exist in the universe. In future work, we plan to study the observational effects of perfect fluid dark matter on the black hole. Furthermore, we plan to study the influence of the perfect fluid dark matter on gravitational lensing phenomena and the evolution of the dark matter in the universe.

\section*{Acknowledgments}
We acknowledge the support from the National Natural Science Foundation of China through grants 11503078, 11573060 and 11661161010. The authors thank M. Kerr for helping improve the manuscript.

\end{document}